# THE PATTERNS OF DIGITAL DECEPTION

GREGORY M. DICKINSON

# THE PATTERNS OF DIGITAL DECEPTION

GREGORY M. DICKINSON[*]

**Abstract:** Current consumer-protection debates focus on the powerful new data-analysis techniques that have disrupted the balance of power between companies and their customers. Online tracking enables sellers to amass troves of historical data, apply machine-learning tools to construct detailed customer profiles, and target those customers with tailored offers that best suit their interests. It is often a win-win. Sellers avoid pumping dud products and consumers see ads for things they actually want to buy. But the same tools are also used for ill—to target vulnerable members of the population with scams specially tailored to prey on their weaknesses. The result has been a dramatic rise in online fraud that disproportionately impacts those least able to bear the loss.

The law's response has been technology centric. Lawmakers race to identify those technologies that drive consumer deception and target them for regulatory restrictions. But that approach comes at a major cost. General-purpose data analysis and communications tools have both desirable and undesirable uses, and uniform restrictions on their use impede the good along with the bad. A superior approach would focus not on the technological tools of deception but on what this Article identifies as the legal patterns of digital deception: those aspects of digital technology that have outflanked the law's existing mechanisms for redressing consumer harm. This Article reorients the discussion from the power of new technologies to the shortcomings in existing regulatory structures that have allowed for their abuse. Focus on these patterns of deception will allow regulators to reallocate resources to offset those shortcomings and thereby enhance efforts to combat online fraud without impeding technological innovation.

## INTRODUCTION

Digital tools have completely revolutionized online commerce.[1] Gone are the days of one-size-fits-all product offerings and static storefronts.[2] Modern

---

[*] Assistant Professor of Law and, by courtesy, Computer Science, the University of Nebraska; Nonresidential Fellow, Stanford Law School, Program in Law, Science & Technology; J.D., Harvard Law School. For their insights and generous comments, thanks to participants at the George Mason University Law & Economics Center Research Roundtable on Regulating Privacy in January 2024 and the Central States Law Schools Association Annual Conference in September 2024. Finally, a special thank you to Rocío Iglesias González for excellent research assistance.

**NOTE:** Because not all platforms support graphics, the figures in this Article are also archived at https://www.bc.edu/content/dam/bc1/schools/law/pdf/law-review-content/BCLR/65-7/dickinson_graphics.pdf [https://perma.cc/Z3L6-YDHN].

online business thrives by understanding and responding to customer needs at an individual level.[3] Using massive databases of historical browsing and shopping data, companies construct detailed consumer profiles,[4] which they then deploy to create online shopping experiences responsive to the preferences of their target audience. Amazon's spot-on product recommendations[5] and the iPhone's effortlessly intuitive user interface[6] are no accidents. Product designs, recommendations, social media activity, and the language, color, layout, and other website and app design features[7] are relentlessly tweaked to create the perfect product and brand experience.[8] Better, more immersive, digital products mean more sales and, just as important, more consumer data, which companies then mine to start the cycle again: creating ever more detailed consumer profiles and enhanced digital marketing campaigns responsive to customers' "senses, emotions, ideas, and behaviors"[9] and then incorporating consumer feedback[10] to develop the next generation of innovative products and services.[11] This cyclical, two-way data flow between consumer and seller means companies know exactly what we want and when we want it, and they will do everything in their power to deliver the products and services we are looking for.

---

[1] *See generally* SHOSHANA ZUBOFF, THE AGE OF SURVEILLANCE CAPITALISM 3–27 (2019); Sunil Erevelles, Nobuyuki Fukawa & Linda Swayne, *Big Data Consumer Analytics and the Transformation of Marketing*, 69 J. BUS. RSCH. 897 (2016); Andrew McAfee & Erik Brynjolfsson, *Big Data: The Management Revolution*, HARV. BUS. REV., Oct. 2012, at 61.

[2] *See* Mohammad Faruk, Mahfuzur Rahman & Shahedul Hasan, *How Digital Marketing Evolved Over Time: A Bibliometric Analysis on Scopus Database*, 7 HELIYON 2021, at 1–2 (documenting this evolution through a review of digital marketing research published between 2000 and 2019).

[3] *See* Baptiste Kotras, *Mass Personalization: Predictive Marketing Algorithms and the Reshaping of Consumer Knowledge*, BIG DATA & SOC'Y, July–Dec. 2020, at 1, 1–2 (discussing modern digital marketing techniques that bring sellers closer to "the old dream of one-to-one, perfectly adjusted selling techniques").

[4] *See id.* at 2 (describing statistical modeling to predict consumer behavior based on large data sets of historical shopping decisions).

[5] *See* Larry Hardesty, *The History of Amazon's Recommendation Algorithm*, AMAZON: SCI. (Nov. 22, 2019), https://www.amazon.science/the-history-of-amazons-recommendation-algorithm [https://perma.cc/5DRF-Q2ET] (providing a brief history of the Amazon product recommendation algorithm's success); Tapan Kumar & Monica Trakru, *The Colossal Impact of Artificial Intelligence in E-Commerce: Statistics and Facts*, 6 INT'L RSCH. J. ENG'G & TECH. 570, 571 (2019) (offering Amazon as a case study of recommendation engines' success and reporting Amazon's algorithm has increased total sales by thirty-five-percent). *But see* Shira Ovide, *Amazon's Open Secret*, N.Y. TIMES (June 18, 2021), https://www.nytimes.com/2021/06/18/technology/amazon-reviews.html [https://perma.cc/969M-7LDA] (describing the problem of bogus online reviews, which can influence Amazon's recommendation system).

[6] *See* Abbas V, *Why Apple Products Feel So Intuitive*, MEDIUM (July 15, 2020), https://medium.com/macoclock/skeuomorphism-the-secret-behind-apples-success-7b7e06348e4c [https://perma.cc/63JE-X2C6] (attributing the success of iPhone user interface in part to its skeuomorphism, the design of its features to "mimic[] the design or feel of their real world counterpart").

[7] Albérico Rosário & Ricardo Raimundo, *Consumer Marketing Strategy and E-Commerce in the Last Decade: A Literature Review*, 16 J. THEORETICAL & APPLIED ELEC. COM. RSCH. 3003, 3011 (2021).

[8] *See* Chen Lou & Quan Xie, *Something Social, Something Entertaining? How Digital Content Marketing Augments Consumer Experience and Brand Loyalty*, 40 INT'L J. ADVERT. 376, 376–78 (2021) (explaining how "branded content marketing affects consumers' brand experience and brand loyalty").

[9] Rosário & Raimundo, *supra* note 7, at 3010.

Digital technologies can be a win for business and consumers both. Sophisticated data-analysis tools "help[] to improve the efficiency, quality, and cost-effectiveness" of products and services provided by a business across nearly every industry.[12] They provide instant, precise information on consumers' evolving needs, leading to increased brand loyalty, perceived value, and consumer satisfaction.[13] Need a gigantic lumbar pillow that looks like a baguette? Amazon knows you do and has them in three sizes.[14] How about bamboo-based toilet paper for a "forest-friendly flush?"[15] It is only a few mouse clicks away, along with anything else your heart could desire. Digital technologies power the most comprehensive and personalized markets for goods and services that the world has ever known.

But digital tools are not limited to the well-behaved. They have also powered a revolution in online fraud.[16] From the Tinder Swindler[17] to identity theft[18] and artificial intelligence (AI)-powered deepfakes,[19] every day seems to bring fresh controversy. Comprehensive consumer databases give online sellers all the information they need either to earn your business or to scam you out of it. Unscrupulous sorts take the latter path: Wonder supplements promising youth, smarts, or bedroom virility in a bottle are the new snake oil; pyramid schemes and stock-picking strategies promise instant riches; and smiling salespersons promise once-in-a-lifetime offers, but only if you click right now.

---

[10] *See* Pratibha A. Dabholkar & Xiaojing Sheng, *Consumer Participation in Using Online Recommendation Agents: Effects on Satisfaction, Trust, and Purchase Intentions*, 32 SERV. INDUS. J. 1433, 1433–34 (2012) (explaining the value of consumer input to product-recommendation algorithms).

[11] Rosário & Raimundo, *supra* note 7, at 3013.

[12] Laith T. Khrais, *Role of Artificial Intelligence in Shaping Consumer Demand in E-Commerce*, 12 FUTURE INTERNET 226, 226 (2020).

[13] Faruk et al., *supra* note 2, at 2.

[14] *See Wepop 40 in 3D Simulation Bread Shape Pillow*, AMAZON, https://www.amazon.com/Simulation-Pillow-Lumbar-Cushion-Stuffed/dp/B07SHP29DM/ [https://perma.cc/7RW5-C5YW].

[15] Brandy, *A Practical Gift Idea for the House: Cloud Paper Sustainable Toilet Paper*, HELLO SUBSCRIPTION, https://hellosubscription.com/2024/01/a-practical-gift-idea-for-the-house-cloud-paper-sustainable-toilet-paper/ [https://perma.cc/5B7B-S9EL] (Jan. 2, 2024).

[16] *See* Jonathan R. Macey, *Fraud in a Land of Plenty*, 118 NW. U. L. REV. 227, 228–30 (2023) (describing the rise in online fraud and suggesting that some attempts to combat it may actually increase fraud by falsely suggesting the problem is under control); Lauren E. Willis, *Deception by Design*, 34 HARV. J.L. & TECH. 115, 123–24, 132 (2020) (detailing the tactics for data collection and analysis that have driven the rise in online fraud); Justin Hurwitz, *Trust and Online Interaction*, 161 U. PA. L. REV. 1579, 1580–81 (2013) (attributing unlawful online behavior in part to the Internet's rapid expansion beyond its small base of early users, who had relied on a "foundation of trust").

[17] *See* THE TINDER SWINDLER (Raw TV, AGC Studios & Gaspin Media 2022) (presenting the story of a con artist who used the dating app to trick a group of women out of thousands of dollars).

[18] *See* Cora Lewis, *Information Theft Is on the Rise. People Are Particularly Vulnerable After Natural Disasters*, ASSOCIATED PRESS, https://apnews.com/article/identity-theft-what-to-do-next-b973836b20e870ddbbb46367a7a18550 [https://perma.cc/EGS6-WUDD] (Sept. 5, 2023) (citing 1.1 million reports of identity theft to the Federal Trade Commission in 2022). *See generally* Lior Jacob Strahilevitz, *Data Security's Unjust Enrichment Theory*, 87 U. CHI. L. REV. 2477, 2477–91 (2020)

Online tricksters are everywhere, and the nation's leading enforcer of consumer-protection laws, the Federal Trade Commission (FTC), has been playing catch-up. Like legitimate products and services, online fraud has become more tailored too. In contrast with the mass emails of old, scammers now stalk and target their victims with expert precision.[20] Often that means schemes targeting vulnerable populations including the elderly, children, non-native English speakers, and military veterans.[21] Digital technologies help scammers sift through potential targets so that their unscrupulous offerings appear on the screens precisely of those individuals most likely to fall victim. Such targeting dramatically increases success rates, and it also decreases the likelihood that the scammers will be found out. The FTC has taken the problem head-on, but it has resources to pursue only a tiny fraction of potential claims—a few dozen out of the hundreds of thousands of online-fraud reports each year.[22] The result has been an explosion of online fraud, with the number of reported instances rising from 181,297 in 2019 to 358,882 in 2023. [23]

---

(surveying the long-standing question of standing for not-yet-harmed victims of identity theft and suggesting unjust enrichment theory as an appropriate basis).

[19] *See* Tiffany Hsu, *As Deepfakes Proliferate, Nations Struggle to React*, N.Y. TIMES, Jan. 22, 2023, at B1 (describing the chaotic response to deepfake technology "that allows people to swap faces, voices and other characteristics to create digital forgeries"); Fed. Trade Comm'n, Comment Letter on Artificial Intelligence and Copyright, No. 2023-6 (Oct. 30, 2023), https://www.ftc.gov/system/files/ftc_gov/pdf/p241200_ftc_comment_to_copyright_office.pdf [https://perma.cc/4D65-STYK] (FTC submission to the U.S. Copyright Office regarding dangers of "generative AI tools," which "turbocharg[e]" deceptive practices).

[20] *See infra* Part I.C; *see also, e.g.*, Joe Hernandez, *That Panicky Call from a Relative? It Could Be a Thief Using a Voice Clone, FTC Warns*, NPR (Mar. 22, 2023), https://www.npr.org/2023/03/22/1165448073/voice-clones-ai-scams-ftc [https://perma.cc/MG5E-UQ9Z] (warning of generative AI's use to target victims with audio clips appearing to come from their relatives).

[21] *See infra* Part I.C.3; James Toomey, *The Age of Fraud*, 60 HARV. J. ON LEGIS. 101, 102–04 (2023) (empirical study assessing the danger of online fraud across age groups).

[22] *See* Chris Jay Hoofnagle, Woodrow Hartzog & Daniel J. Solove, *The FTC Can Rise to the Privacy Challenge, but Not Without Help from Congress*, BROOKINGS INST. (Aug. 8, 2019), https://www.brookings.edu/articles/the-ftc-can-rise-to-the-privacy-challenge-but-not-without-help-from-congress/ [https://perma.cc/N2HL-7BWB] (noting that "[r]esources are the FTC's greatest constraint" and that FTC attorneys must "weigh[] decisions on where to target limited enforcement resources" because "[t]he FTC can only bring actions against a small fraction of infringers").

[23] FED. TRADE COMM'N, CONSUMER SENTINEL DATA BOOK 2021, at 84 (2022), https://www.ftc.gov/system/files/ftc_gov/pdf/CSN%20Annual%20Data%20Book%202021%20Final%20PDF.pdf [https://perma.cc/DA5D-2ZYB] [hereinafter CONSUMER SENTINEL DATA BOOK 2021] (providing the

Seeing so many Americans affected, and with even more powerful tools on the horizon, legislatures and regulators have sprung to action. To bolster the FTC's traditional, case-by-case approach to combating unfair competition, lawmakers have proposed (and in some instances enacted) new statutes and regulations to restrict the digital technologies that power online deception.[24] The idea is to preserve the FTC's scarce enforcement resources by enacting prophylactic restrictions on the technologies that drive deception instead of waiting to pursue wrongdoers after the fact.

This Article warns that that approach is a mistake for two reasons. First, what is new and dangerous about technology-powered scams is not any special power to deceive but their unprecedented efficiency.[25] Digital marketing technologies bring the marginal cost of scamming one more consumer near to zero. What, after all, is the cost of one more targeted ad? Low-cost schemes are nothing new, and neither are targeted cons. What is different about modern online fraud is that it is both cheap *and* highly personalized. Whereas once these scams might have required a skilled con artist to invest hours of her time locating a potential victim and learning her vulnerabilities, personalized schemes now can be designed, targeted, and deployed on a massive scale and at far lower cost. This observation has important implications for how the law should respond to the threat. It suggests that online deception demands no new *substantive* law, for it already fits comfortably within the existing regulatory regime, but that low-cost digital deception has outstripped existing enforcement resources.

Second, although across-the-board restrictions on digital technologies might have some effect on online fraud, they would do so only at a major cost to innovation.[26] Digital tools have revolutionized not just online fraud, but the entire market. Restricting those tools will have consequences far beyond internet scammers, to the detriment of software and services used by consumers around the country every day. Across-the-board regulation of key technologies

---

number of online-fraud reports made in 2019); Fed. Trade Comm'n, Consumer Sentinel Network, *Fraud and ID Theft Maps*, TABLEAU PUB., https://public.tableau.com/app/profile/federal.trade.commission/viz/FraudandIDTheftMaps/AllReportsbyState [https://perma.cc/59GB-3R47] (July 24, 2024) [hereinafter *Fraud and ID Theft Maps*] (interactive report of fraud claims by time period and category, updated regularly); *see* FED. TRADE COMM'N, CONGRESSIONAL BUDGET JUSTIFICATION FISCAL YEAR 2024, at 12, 15–24 (2023), https://www.ftc.gov/system/files/ftc_gov/pdf/p859900fy24cbj.pdf [https://perma.cc/6CLJ-ABPW] [hereinafter CONGRESSIONAL BUDGET JUSTIFICATION FISCAL YEAR 2024] (justifying an increased budget in part to combat "frauds using new technologies in the areas of online and mobile transactions" and highlighting the FTC's myriad enforcement actions, many of which have been against online fraud).

[24] *See infra* Part II.A (collecting numerous bills and regulations proposed to combat online deception).

[25] *See infra* Part I.C.

[26] *See infra* Part II.B.

would increase costs and reduce product quality for everyone, for a comparatively minor benefit: scammers would be forced to adopt new tools or, more likely, to ignore the restrictions altogether.

Instead of enacting new technology restrictions, this Article argues, regulators should bolster enforcement efforts in a different way—by coordinating governmental enforcement efforts with those of private litigants.[27] Deceptive online practices are prohibited not only by the federal FTC Act, but also by state consumer-protection laws which, unlike the FTC Act, provide victims with a private right of action.[28] Unfortunately, however, procedural shortcomings in the law have enabled many online scammers to perpetuate their schemes without facing the private lawsuits that would ideally provide a check against such wrongdoing.

In particular, four types of online schemes—what this Article identifies as the patterns of deception—have been especially resistant to private enforcement efforts:[29] (1) *fly-by-nighters*, whose highly mobile operations or location in foreign jurisdictions makes private enforcement difficult; (2) *nickel-and-dimers*, who operate at a large scale but extract small sums of money from people who individually lack sufficient interest to pursue litigation; (3) *user-interface shapeshifters*, whose varied and quickly changing user interfaces pose an obstacle to aggregate litigation; and (4) *calculated arbitrators*, whose terms of service include agreements requiring individualized arbitration of claims and barring consumers from seeking class relief.

Finally, taking in hand this taxonomy of digital ne'er-do-wells, this Article proposes a coordinated public-private response to digital deception that relies on private litigation supplemented by federal enforcement resources strategically redeployed to combat those deceptive operations that are most resistant to private enforcement. Focus on these legal patterns of deception will offset the procedural limitations of private litigation, thereby enhancing the overall effectiveness of efforts to combat online fraud, while avoiding the impediments to technological innovation that would come from across-the-board technology restrictions.

---

[27] *See infra* Part III.
[28] *See* Allan Bruce Currie, *A Private Right of Action Under Section Five of the Federal Trade Commission Act*, 22 HASTINGS L.J. 1268, 1268 (1971) (noting that "federal courts have held that there is no private right of action under [Section 5 of the FTC Act], declaring that only the [FTC] could institute an action for its violation"); *infra* notes 139–143 and accompanying text (discussing state legislative efforts in response to dark patterns).
[29] *See infra* Part III.B.

## I. THE NEW ERA OF DECEPTION

Digital marketing technologies may be new, but commercial deception certainly is not. This Part explores what is new and what is not about online fraud, setting the stage for a discussion, in Part II, of legislative and regulatory efforts to combat it. Section A of this Part gives a brief history of commercial fraud and the law's static response. Sections B and C discuss the tools involved in the shift to digital markets and their potential for misuse by scammers.

### A. Fraud: Tale as Old as Time

If there are any things new under the sun, human trickery is not among them. From Homer, we have Odysseus introducing himself to Polyphemus as "Nobody—that's my name" so that the cyclops's cry that "Nobody's killing me now" would go unanswered.[30] And Genesis gives us Jacob, who, dressed literally in goats' clothing, impersonated his hairier brother, Esau, to obtain his father's blessing.[31] The practice is not limited to heroes and patriarchs. It is an ineluctable lesson of history that everyone—from kings and queens to ordinary folk—is apt to behave badly when money or power is at stake.

The path of our common law tells the same story. The cases document recurring scuffles between buyers and sellers as each tries to win for herself the best possible bargain, sometimes stretching beyond the confines of the law to do so. Early English cases include a textile buyer displeased when his seller delivered only forty-six instead of the promised fifty sacks of wool;[32] a merchant who sold a cask of malmsey wine "knowing the wine to be stale" yet nonetheless "warranted the same to be good and drinkable";[33] a seller who added water to dried hops sold by weight, hoping to increase their price;[34] and, in the domestic realm, a scheming, would-be brother-in-law who "placed [the plaintiff] in a certain bed until [the plaintiff] was asleep," led his (that is, the defendant's) sister there and "placed her in the said bed naked with [the plaintiff]" and then brought two witnesses into the room to observe them "lying to-

---

[30] HOMER, THE ODYSSEY 9.410–55, pp. 223–24 (Robert Fagles trans., Viking Penguin 1996); *cf.* PLATO, *Protagoras* 313d, p. 751 (Stanley Lombardo & Karen Bell trans.) (analogizing sophists to "merchants who market food" and "don't know what is good or bad for the body—they just recommend everything they sell"), *in* PLATO COMPLETE WORKS (John M. Cooper ed., Hackett 1997).

[31] *Genesis* 27:1–45.

[32] *See* Drew Barantine's Case, YB Mich. 13 Hen. IV, fo. 1, pl. 4 (1411) *reprinted in* JOHN BAKER, BAKER AND MILSOM SOURCES OF ENGLISH LEGAL HISTORY: PRIVATE LAW TO 1750, at 561 (2d. ed. 2010) [hereinafter BAKER & MILSOM].

[33] *See* Anon. (1491) Caryll's Reports, Seldon Soc. Vol. 115, 73 *reprinted in* BAKER & MILSOM, *supra* note 32, at 566.

[34] *See* Parkinson v. Lee, 102 Eng. Rep. 389, 389 (KB 1802).

gether alone and naked" in an effort to force marriage on the plaintiff against his will.[35]

The case law of industrializing America shows technological advancement, but no change in behavior. In Massachusetts, a seller misrepresented the condition of a steam engine, which he claimed to be "a twenty-horse power engine" "fit for mining purposes," "free from rust," and "standing but two or three years."[36] In North Carolina, a seller mixed bales of cotton with sand to increase their weight,[37] and a lender induced a drunken man to enter a contract to repay another's debt.[38] And in New York, a man found a lost herd of twenty-one sheep by the roadside, inquired as to who had lost them and then, pretending not to know where they were located, arranged to buy the missing herd at a discount as a "favor" to the man who had lost them.[39] In short, times have changed, but humans have not. From medieval merchants, con artists of early America, and door-to-door encyclopedia salesmen in the twentieth century,[40] sellers will pitch their products in the very best light, sometimes bending or breaking the truth as they do so.

Across the centuries, the law's response has remained constant, too. Despite dramatic social and technological change, the legal guideposts remain the same:[41] Sellers are held to a standard of honesty and fair dealing and must not deceive their customers,[42] while buyers have a complementary obligation to look out for their own interests; they cannot complain afterward about problems that should have been obvious at the time of the purchase.[43] That the law of fraud should remain static in the face of technological and social change is notable, but unsurprising, for what is wrong about fraud "is not the particular instrument chosen either to deceive or coerce . . . but that such deception and coercion compromised the autonomy of the [plaintiff] in the first place."[44] Indeed, turning to the present day, the basic common-law principles remain undisturbed even following Congress's move in the early twentieth century to nationalize consumer-protection law and vest enforcement authority in the FTC.[45] Like the common law before, modern consumer-protection law under the FTC Act prohibits all manner of "deceptive acts or practices in or affecting commerce,"[46] but only if they are "likely to mislead the consumer acting reasonably in the circumstances."[47] Consumers still have an obligation to look after their own interests. If a "few misguided souls" make bakery purchases falsely believing that "all 'Danish pastry' is made in Denmark," they are out of luck.[48]

---

[35] Trote v. Lynet (1358), *reprinted in* BAKER & MILSOM, *supra* note 32, at 556.

[36] Hazard v. Irwin, 35 Mass (18 Pick.) 95, 95 (1836); *see also, e.g.*, Carondelet Iron Works v. Moore, 78 Ill. 65, 66 (1875) (inferior white-mottled iron sold as higher quality mill iron).

[37] Stout v. Harper, 51 N.C. (6 Jones) 347, 347–48 (1859).

[38] Guy v. McLean, 12 N.C. (1 Dev.) 46, 46 (1826).

[39] Bench v. Sheldon, 14 Barb. 66, 66–67 (N.Y. Gen. Term. 1852).

[40] *See* Laura Barnett, *Death of a Salesman: No More Door-to-Door Britannica*, THE GUARDIAN (Mar. 14, 2012), https://www.theguardian.com/media/shortcuts/2012/mar/14/britannica-death-salesmen-door [https://perma.cc/93G3-33R9] (detailing the historical phenomenon of door-to-door encyclopedia salesmen and bidding them farewell as Britannica stops printing encyclopedias); *see also* FTC Cooling-Off Rule, 16 C.F.R. §§ 429.0–.1 (2024) (giving buyers three business days to cancel certain purchases from high-pressure, door-to-door sellers).

[41] *See* Paula J. Dalley, *The Law of Deceit, 1790–1860: Continuity Amidst Change*, 39 AM. J. LEGAL HIST. 405, 419–41 (1995) (reviewing early English and American cases, identifying "[h]onesty and fair dealing" and "prudent attention to one's own affairs" as the leading policy concerns of fraud and contract law, and observing that courts of nineteenth-century America "developed, but did not fundamentally change" the law inherited from England). *See generally* Richard A. Epstein, *The Static Conception of the Common Law*, 9 J. LEGAL STUD. 253, 253, 256 (1980) (reasoning that "the importance attributable to changing social conditions as a justification of new legal doctrines is overstated and quite often mischievous" and offering fraud as an example of where the law has and should remain constant).

[42] *See, e.g.*, Kenrick v. Burges, Moo. 126, pl. 273 (QB 1583) (recognizing viability of fraud action against a woman who, after the death of her husband, purported to sell a leasehold interest in the property to a third party despite "knowing that she had no right to the term"), *reprinted in* BAKER & MILSOM, *supra* note 32, at 568; Mitchell v. Christensen, 31 P.3d 572, 574–76 (Utah 2001) (recognizing a fraud claim for seller's failure to disclose known leaks in the backyard swimming pool); Booker

### B. Digital Hucksters

If the battle between consumers and sellers is so ancient, why all the fuss about digital deception? Are not online hucksters just the newest players in the same old story? Yes, and no. Much digital deception is nothing more than a direct translation of age-old tactics to the online world. Both sides of the tussle

---

T. Washington Constr. & Design Co. v. Huntington Urb. Renewal Auth., 383 S.E.2d 41, 44–46 (W. Va. 1989) (breach of covenant of seisin where a fee simple interest was purportedly conveyed but the seller knew it possessed only a life estate).

[43] *See, e.g.*, Lopus v. Chandler (no. 4) (1606) (reasoning that if someone negotiates in person "to buy my horse as it stands in the stable . . . and I say that it ambles, whereas in truth it trots," he will not have an action "for it was his own overcredulity which deceived him"), *reprinted in* BAKER & MILSOM, *supra* note 32, at 570–74; Spreitzer v. Hawkeye State Bank, 779 N.W.2d 726, 737 (Iowa 2009) (explaining that fraud plaintiffs cannot "blindly rely on a representation" but must "utilize their abilities to observe the obvious").

[44] Epstein, *supra* note 41, at 256.

[45] *See* Federal Trade Commission (FTC) Act, ch. 311, 38 Stat. 717 (1914) (codified as amended at 15 U.S.C. §§ 41–58) (creating the FTC and defining its powers and duties).

[46] FTC Act § 5(a); 15 U.S.C. § 45(a)(1).

[47] Letter from James C. Miller III, Chairman, Fed. Trade Comm'n, to Representative John D. Dingell, Chairman, Comm. on Energy and Com., FTC Policy Statement on Deception (Oct. 14, 1983), https://www.ftc.gov/system/files/documents/public_statements/410531/831014deceptionstmt.pdf [https://perma.cc/7YFJ-2LLF] [hereinafter FTC Policy Statement on Deception]; *cf.* 15 U.S.C. § 45(n) (limiting FTC's enforcement authority over "unfair" practices to those "not reasonably avoidable by consumers themselves").

[48] Heinz W. Kirchner, 63 F.T.C. 1282 (1963) (offering the Danish pastry example and concluding that "[a] representation does not become 'false and deceptive' merely because it will be unreasonably misunderstood" by a small number of persons).

remain the same, just on a new playing field. But some features of the digital marketplace make consumers unusually vulnerable.

With the digital age have come new modes of doing business. Data is the driver of business, and the spoils go to those who harness it best. Most obvious are the Facebooks, Googles, and Netflixes of the world, whose very business is data. What use are social networks, cloud services, or streaming platforms, after all, without the countless gigabytes of data that power them? Such entities are attractive to consumers in part *because of* their size, enabling them to attract ever-larger numbers of users and data.[49] Why join a newly launched social network with few users when you could instead join Facebook or LinkedIn where all of your friends and colleagues already have accounts? The answer is that you do not, at least until some competitor offers a revolutionary product or seamless onramp.[50] For the tech industry, this has meant that a small number of massive companies dominate important segments of the industry and hold a treasure trove of data about their users.

Yet the shift toward data-driven products and services has effects far beyond the dominant tech firms. Sure, they may have the very most data, but everyone is getting in on the game, from fresh tech startups to brick-and-mortar stalwarts like Walmart and Ford.[51] Everywhere one turns, a new app or website demands an email address and phone number before allowing its users even to open the app or browse merchandise. The boldest such apps even request access to users' smartphone contacts and location data. These data requests are not mere nosiness (although they are certainly that too), but the lifeblood of modern business. With users' email addresses, sellers can circulate ads, promote sales, and build brand loyalty;[52] with phone numbers, which rarely change, sellers can track users' shopping habits across websites and learn what they are currently interested in purchasing and at what prices;[53] and with location data and contacts they can determine when users shop at their stores or their competitors and identify other, similar consumers who may have the same shopping habits.[54] Data like this are simply too powerful to ignore. Even sellers who do not directly collect such data use it indirectly by relying on targeted advertising tools offered by data-gathering titans like Google and Facebook.[55]

---

[49] *See* CARL SHAPIRO & HAL R. VARIAN, INFORMATION RULES: A STRATEGIC GUIDE TO THE NETWORK ECONOMY 179–84 (1999) (observing this phenomenon and explaining that user growth fueled by demand-side economies of scale produces "especially strong positive feedback").

[50] *See id.* at 190–96 (offering evolution and revolution as the "two basic approaches for dealing with the problem of consumer inertia"); *see also* Gregory M. Dickinson, *Big Tech's Tightening Grip on Internet Speech*, 55 IND. L. REV. 101, 105–11 (2022) (discussing the concern that social media platforms' size may impede competition).

[51] *See* Mary E. Morrison, *Ford Leverages Data, Analytics to Drive CX Transformation*, WALL ST. J. (July 23, 2018), https://deloitte.wsj.com/articles/ford-leverages-data-analytics-to-drive-cx-transformation-1532319386 [https://perma.cc/5HJR-JYHF] (discussing how Ford uses consumer data to improve customer experience with complex new vehicles); Bernard Marr, *Really Big Data at Walmart: Real-Time Insights from Their 40+ Petabyte Data Cloud*, FORBES, https://www.forbes.com/sites/bernardmarr/2017/01/23/really-big-data-at-walmart-real-time-insights-from-their-40-petabyte-data-cloud/ [https://perma.cc/E6T3-TRDX] (May 13, 2019) (describing Walmart's collection and processing of 2.5 petabytes of data from their customers each hour); Joann Muller, *How Ford Is Using Big Data to Change the Way We Use Our Cars*, FORBES (Oct. 22, 2015), https://www.forbes.com/sites/joannmuller/2015/10/22/how-ford-is-using-big-data-to-change-the-way-we-use-our-cars/ [https://perma.cc/V6GS-KKHP] (discussing Ford's collection of customer data).

Data collection is only the beginning. After gathering customers' digital identifiers and basic demographic information such as age, gender, and zip code, companies use those data to build individualized profiles to help predict their interests, hobbies, and likely shopping habits.[56] Suppose a married, twenty-seven-year-old woman recently moved into a new house. Home-goods retailer Beds Baths and Babies knows this because she updated her online account and provided a new shipping address. The retailer's real-estate database shows the address to be a four-bedroom house in a posh neighborhood (in contrast with her previous address, a one-bedroom flat in a budget-friendly apartment complex). The store sends her an email inviting her to download its app, which will provide access to a reliable stream of twenty-percent-off coupons. She installs the app and logs in with her smartphone, excited at her prospective savings. In exchange, the store now has access to her phone number and mobile device identifier and can associate her Beds Baths and Babies account with her internet-search and online-shopping history. Connecting the dots between the woman's new home and her recent searches for luxury all-terrain strollers, Beds Baths and Babies realizes it has hit the jackpot—a customer at the center of its target demographic and likely soon to make numerous purchasing decisions.

---

[52] *See* Amy Nichol Smith & Kelly Main, *Best Email Marketing Software & Tools*, FORBES, https://www.forbes.com/uk/advisor/business/software/best-email-marketing-software/ [https://perma.cc/AD73-ZGMF] (Oct. 17, 2023) (noting that "[e]mail marketing remains one of the most effective tools for any company to reach engaged customers").

[53] *See* Scott Rosenberg, *Phone Numbers Are the New Social Security Numbers*, AXIOS (Mar. 5, 2019), https://www.axios.com/2019/03/05/phone-numbers-are-the-new-social-security-numbers-1551732831 [https://perma.cc/ZY9L-EXXJ] (describing companies' reliance on phone numbers to track customers' habits and suggest relevant products).

[54] *See* Fareena Sultan & Syagnik Banerjee, *Enhancing Customer Insights with Public Location Data*, HARV. BUS. REV. (June 12, 2018), https://hbr.org/2018/06/enhancing-customer-insights-with-public-location-data [https://perma.cc/L85M-NAGQ] (detailing companies' use of consumer location data to predict purchasing behavior).

[55] *See infra* note 75 and accompanying text (explaining Google's and Facebook's businesses of selling targeting ads).

[56] *See* Karen Yeung, *Hypernudge: Big Data as a Mode of Regulation by Design*, 20 INFO., COMMC'N & SOC'Y 118, 118–19 (2017) (explaining how online sellers build extensive user profiles to predict customer preferences and customize their shopping experiences); FRANCISCO LUPIÁÑEZ-VILLANUEVA ET AL., DIRECTORATE-GENERAL FOR JUST. & CONSUMERS, BEHAVIOURAL STUDY ON UNFAIR COMMERCIAL PRACTICES IN THE DIGITAL ENVIRONMENT 19–22 (2022) (describing online sellers' profiling strategies and briefly summarizing the academic literature).

Now when this customer visits the Beds Baths and Babies website or opens its app, she will be greeted by color-popping ads for baby bottles, cribs, and swaddle blankets. Her email address will be added to the company's targeted marketing campaign for parents to be, assuring regular emails promising to help get her new home "baby ready." And because Beds Baths and Babies has data-sharing arrangements with commercial data brokers who resell data about its customers' shopping habits to other companies, baby-themed marketing will follow her all over the internet. Ads full of smiling babies decked out in cutesy outfits will now adorn all of her social media feeds and search results.

Still, that is not the end of the data's story. The customer's shopping history will be aggregated with data about similar customers at Beds Baths and Babies and elsewhere and "mined" for new insights into the habits and preferences of new mothers, which will help the company better target future customers like her. Naturally such analysis would show that expecting mothers buy cribs and strollers. Any salesperson could tell you that. What sets modern big data and machine learning approaches apart from their analogue predecessors is the ability to discover complex or counterintuitive patterns in data sets—correlations that could not have been discovered by ordinary human intuition or analysis.[57] For example, supermarkets famously (and, sadly, apocryphally)[58] used data mining to discover that men who buy diapers also buy beer and so began placing the two together on store shelves. In our hypothetical customer's case, the analysis will reveal what snack foods and exercise equipment are most appealing to pregnant, mid-twenties women in her geographic region and at her estimated income level; what style and color automobiles they are most likely to purchase; and even such things as which of several potential slogans advertising life insurance are most likely to get a click from her demographic.

That last data element, the relative attractiveness to a particular demographic of various advertising strategies, is the product of A/B testing, a common technique by which companies tweak app designs by splitting users into groups and comparing usage metrics between groups.[59] Such testing is a mainstay of app and web design because it helps companies optimize the experiences of their users. For example, a company might split users into A and B test groups, present each with a slightly different user interface, and measure which group is able to complete routine tasks more quickly or rates the app more highly in the app store.

---

[57] *See* Yeung, *supra* note 56, at 119 (observing that "[a] key contribution of Big Data is the ability to find useful correlations within data sets *not capable of analysis by ordinary human assessment*").

[58] *See Diaper-Beer Syndrome*, FORBES, https://www.forbes.com/forbes/1998/0406/6107128a.html [https://perma.cc/63PH-F9JB] (June 6, 2013) (recounting the legend of the company that used data mining to discover that men who bought diapers at a grocery store also would pick up beer while already there).

The same technique can be used to make innumerable tweaks to the "choice architecture"[60] an app or website presents to its users—for example, where buttons are placed, how questions are phrased, or whether to use an opt-in or opt-out system for email marketing. Of course, there is no need to run A/B tests to know that companies will prefer an opt-out system, that is, one where a customer's inaction leads to enrollment, over an opt-in system, where inaction goes the other way.[61] But what about when a company is deciding which products to list most prominently in search results or how to arrange the tree of menu options in its app? There, A/B testing is indispensable, and those who master it stand to reap huge rewards.[62]

Comparing groups by differences in outcome is nothing new. Sellers have done that for ages, either intuitively or, for larger companies, with focus groups, surveys, and advertising data. What is different in the digital context is the scale and ease of such studies. Focus groups and surveys are expensive,[63] and sellers can only try so many iterations. A/B testing for websites and apps is different. Users can be split into thousands of groups, each presented with slightly different versions of a user interface or ad, and the process iterated until the very most effective design[64] is identified to sell the most widgets or enroll the most subscribers.[65] There is no need to manually apportion users into groups, choose which group gets which interface, or even design the interface variations to be tested. The whole process can be overseen by computer algorithm.[66]

---

[59] For a general overview of A/B Testing, see generally Amy Gallo, *A Refresher on A/B Testing*, HARV. BUS. REV. (June 28, 2017), https://hbr.org/2017/06/a-refresher-on-ab-testing [https://perma.cc/HR4P-GWA8] (explaining what A/B testing is, how it works, how companies use it, and the common mistakes while using it); DAN SIROKER & PETE KOOMEN, A/B TESTING: THE MOST POWERFUL WAY TO TURN CLICKS INTO CUSTOMERS (2013).

[60] Choice architecture refers to the context in which a decision is to be made—in what words the options will be presented, how many choices offered, what the paper form or digital interface that conveys those choices will look like, and so forth. Like physical-world architecture, the task is unavoidable; a building must take *some* shape, and so must decision-making contexts. *See* RICHARD H. THALER & CASS R. SUNSTEIN, NUDGE: THE FINAL EDITION 3–5, 16–18 (2021) (explaining the concept of choice architecture and providing examples); *cf.* LAWRENCE LESSIG, CODE AND OTHER LAWS OF CYBERSPACE 30–42 (1999) (considering the important differences between physical-world architectures and the code-based architectures of the digital world).

[61] Whether and under what circumstances an individual's failure to opt-out of direct marketing will constitute valid consent under European Union data privacy laws is a matter of ongoing dispute. *See* Dan Milmo, *Meta Dealt Blow by EU Ruling That Could Result in Data Use 'Opt-in,'* THE GUARDIAN (Jan. 4, 2023), https://www.theguardian.com/technology/2023/jan/04/meta-dealt-blow-eu-ruling-data-opt-in-facebook-instagram-ads [https://perma.cc/25B2-6TAX] (reporting on ruling by Ireland's Data Protection Commission that Facebook users must opt in to having data used for targeted ads); Angelique Chrisafis, *France Cracks Down on Junk Mail with Trial Opt-in System*, THE GUARDIAN (Sept. 1, 2022), https://www.theguardian.com/world/2022/sep/01/france-cracks-down-on-junk-mail-with-trial-opt-in-system [https://perma.cc/GF46-D65B] (explaining a trial system in France in which residents may opt to put a "yes to advertising" sticker on their mailbox to receive unaddressed advertising mail).

[62] *See generally* Leo Kelion, *Why Amazon Knows So Much About You*, BBC, https://www.bbc.co.uk/news/extra/CLQYZENMBI/amazon-data [https://perma.cc/J97X-4K45] (explaining how Amazon collects its customers' data).

Digital-world choice architectures are easier to optimize for a second reason, too: They are far easier to change. Recall the grocery store selling diapers with beer. Iterative A/B testing for physical-world product placements like that one would require endless rearranging of aisles, shelves, and store layouts. Major changes, perhaps the addition of a walk-in, refrigerated beer cave, might even require construction work and caution tape. Online choice architectures, by contrast, can be rearranged on the fly, with only a few lines of code and little disruption to users. Indeed, they could be rearranged for every single user, so that each user explores the shop with virtual aisles arranged to suit her own preferences.

Modern data collection and analysis techniques have shifted the balance of power between buyers and sellers. Massive troves of historical shopping and browsing data mean that sellers often know what their customers want better than the customers themselves. Adding A/B testing and on-the-fly tweaks to digital choice architectures, companies can present customers offers for the exact products they will want, in maximally attractive terms, customized even for the individual shopper.

---

[63] App makers and websites, by contrast, typically run A/B tests on their current user base. Indeed, they need not even ask permission, making for more representative sample groups. The proposed federal Deceptive Experiences to Online Users Reduction (DETOUR) Act would have imposed notice and consent requirements on any "behavioral or psychological experiment or research" by an online service of its users, including A/B testing. *See* Deceptive Experiences to Online Users Reduction (DETOUR) Act, H.R. 6083, 117th Cong. § 3(a) (2021).

[64] Mechanized A/B testing for "effective" designs that fails to consider why they are so effective can lead to misleading ad phrasings or interface designs. *See, e.g.*, Compl. at 4–5, Credit Karma, LLC, No. C-4781 (F.T.C. Jan. 19, 2023) (asserting that A/B testing led the credit card company to choose ad reading "You're pre-approved" over one informing prospective customers of "Excellent" odds of approval even though the company did not actually preapprove consumers and nearly one-third of those who clicked the ad and applied were ultimately denied credit).

[65] *See* James Grimmelmann, *The Law and Ethics of Experiments on Social Media Users*, 13 COLO. TECH. L.J. 219, 233–35 (2015) (describing the "nearly constant testing" app makers engage in to improve their products and the ethical implications of experimenting on human users without consent); Ron Kohavi & Stefan Thomke, *The Surprising Power of Online Experiments*, HARV. BUS. REV., Sept.–Oct. 2017, at 74, 74 (describing the massive scope and revenue implications of A/B testing at Microsoft and other leading technology companies).

[66] *See generally* Iavor Bojinov & Somit Gupta, *Online Experimentation: Benefits, Operational and Methodological Challenges, and Scaling Guide*, HARV. DATA SCI. REV., Summer 2022, at 2 (providing an A/B testing primer and "explaining the benefits, challenges . . . and best practices in creating and scaling" the experimentation).

### C. Personalization at Scale

All of this may be startling, but it is very often a good thing. Sellers avoid pumping dud products no one wants, consumers get to browse products customized to their own likely preferences, instead of what everyone else is buying, and they see ads for products they might actually want to buy. Unfortunately, the same tools can also be misused.

1. Consumer Data

Just as machine learning algorithms could mine consumer data to reveal that a mother is expecting a child and more likely to be interested in purchasing life insurance, the same techniques might reveal that she is unusually gullible, irrationally risk averse, or prone to making mistakes when parsing written English.[67] Instead of targeting her with useful product offers, a seller might then target her for deceptive offers designed to trick her out of her money instead of earning her business. The deceptive offers would be all the more alluring because, unlike traditional marketing directed toward the masses, she could be targeted with individual schemes to prey on her unique needs and vulnerabilities. That is exactly what has happened over the last several decades as advances in data collection, storage, and analysis have been applied to create both hyper-intuitive and targeted product designs and also, what is the dark side of the same technologies, astoundingly effective consumer fraud schemes.[68]

Businesses can now rely on data brokers to access vast troves of consumer data to help them target a consumer's particular vulnerabilities. Data brokers collect and store data on almost every U.S. household and every commercial transaction.[69] They aggregate information on individuals from their public so-

---

[67] For details regarding commonly collected categories of consumer data and their use in targeted advertising, see generally Sophie C. Boerman, Sanne Kruikemeier & Frederik J. Zuiderveen Borgesius, *Online Behavioral Advertising: A Literature Review and Research Agenda*, 46 J. ADVERT. 363 (2017); FED. TRADE COMM'N, A LOOK BEHIND THE SCREENS: EXAMINING THE DATA PRACTICES OF SOCIAL MEDIA AND VIDEO STREAMING SERVICES (2024), https://www.ftc.gov/reports/look-behind-screens-examining-data-practices-social-media-video-streaming-services [https://perma.cc/BQQ7-6WNF].

[68] *See infra* Part I.C.

[69] FED. TRADE COMM'N, DATA BROKERS: A CALL FOR TRANSPARENCY AND ACCOUNTABILITY 1–3 (2014), https://www.ftc.gov/system/files/documents/reports/data-brokers-call-transparency-accountability-report-federal-trade-commission-may-2014/140527databrokerreport.pdf [https://perma.cc/4WQG-JDUG].

cial media posts, online shopping history, browsing history, real estate records, voter registration information, marriage, divorce, and birth records, and other sources to create consumer profiles of nearly every person in the United States.[70] The profiles flag attributes of interest to sellers, such as the individual's personal characteristics and shopping and social habits. For example, an individual's profile might include the following:

| | |
|---|---|
| ☐ Age 68 | ☐ Business Professional |
| ☐ White | ☐ Affluent Baby Boomer |
| ☐ Male | ☐ Hunting & Shooting |
| ☐ Republican | ☐ Allergy Sufferer |
| ☐ 3 Bedroom Home | ☐ Senior Products Buyer |
| ☐ Net Worth > $1M | ☐ Christian |
| ☐ Married | ☐ Plus-Size Apparel |
| ☐ Investor | ☐ Gambling |

A real consumer profile would include many more attributes and also the raw data that produced them. With these attributes in place, the consumer and her purchase history can then be compared with other consumers in the data broker's database to make predictions about what the consumer is interested in purchasing right now and what types of products she is likely to purchase in the future.[71]

Sellers who wish to make use of such data can do so in several ways. Those without established businesses and customer databases, or those who wish to expand them, can purchase databases of prospective customers from data brokers based on the customers' personal characteristics, likelihood of purchasing particular types of products, or both.[72] Alternatively, sellers who already have contact information for past or prospective customers might instead coordinate with a data broker to augment their existing databases by adding additional information that the data broker holds on those individuals.[73] For example, a seller who has only a list of names and email addresses might work with a data broker to obtain information about those individuals' net worth, geographic region, or propensity to purchase particular products. Finally, perhaps most importantly,[74] sellers can take advantage of such information without handling consumer data directly at all. Google, Facebook, and other entities offer targeted advertising tools that allow sellers to purchase ad space targeted exclusively to consumers who fall within desired categories.[75]

---

[70] *See id.* at 11–18 (surveying data brokers' sources of data and collection methodologies).
[71] *Id.* at 1–19.
[72] *Id.* at 23–25.
[73] *Id.*
[74] More than one-quarter of those who reported losing money to fraud in 2021 said the scheme started on social media. *Social Media a Gold Mine for Scammers in 2021*, FED. TRADE COMM'N: DATA SPOTLIGHT BLOG 1 (Jan. 2022), https://www.ftc.gov/system/files/attachments/blog_posts/Social%20media%20a%20gold%20mine%20for%20scammers%20in%202021/social_media_

2. Tailored Scams

Data-driven advertising campaigns can be fine-tuned to such precise audiences that they have become a powerful tool for online trickery.[76] For example, once a website learns which item a customer is shopping for (either using its own data or that provided by a partner), a common tactic is to advertise a fake, limited-time sale for the specific item that the customer is currently shopping for.[77] In truth the item is always available at the specified price, but by recharacterizing the offer as a limited-time sale, the site creates a false sense of urgency hoping to spur a sale even when the customer might otherwise have waited to research competing products and sellers. To increase pressure and augment the technique's power, the site might also include a fake countdown timer that ominously ticks off the seconds remaining on the offer.[78] Or, instead of counting off the seconds, the site might include a fake inventory count proclaiming, "Buy now, only 5 left in stock!"[79]

---

spotlight.pdf [https://perma.cc/D5TP-858J] (noting that "scammers c[an] easily . . . target people with bogus ads based on personal details such as their age, interests, or past purchases").

[75] *See* Megan Graham & Jennifer Elias, *How Google's $150 Billion Advertising Business Works*, CNBC, https://www.cnbc.com/2021/05/18/how-does-google-make-money-advertising-business-breakdown-.html [https://perma.cc/2XVP-SFQC] (Oct. 13, 2021) (discussing workings of Google's advertising business); *How Do Facebook Ads Target You?*, CBS NEWS (Apr. 13, 2018), https://www.cbsnews.com/news/how-do-facebook-ads-target-you/ [https://perma.cc/7KTG-5URY] (describing various ways advertisers use Facebook to target consumers).

[76] *See* Trade Regulation Rule on Commercial Surveillance and Data Security, 87 Fed. Reg. 51273, 51274, 51290 (Aug. 22, 2022) (observing that "[w]hile, in theory . . . personalization practices have the potential to benefit consumers . . . they have facilitated consumer harms that can be difficult if not impossible for any one person to avoid" and proposing Magnuson-Moss rulemaking to prohibit harmful consumer surveillance and targeting practices).

[77] Jaclyn Peiser, *Black Friday Steals Might Not Be So. Avoid Deceptive Pricing with These Tips*, WASH. POST, Nov. 22, 2023, at A14 (describing pricing technique of raising prices "only to mark it down to the original price while marketing it as a limited-time, steep discount").

[78] *See, e.g.*, *Boohoo 'Broke Advertising Rules,' BBC Watchdog Finds*, BBC (Dec. 4, 2018), https://www.bbc.com/news/business-46441526 [https://perma.cc/Q7JE-F4CH] (reporting fast-fashion retailer Boohoo's use of fake countdown timers that restarted once the timer reached zero).

[79] *See* Herb Weisbaum, *The Travel Website You're Using Says There's 'Only 1 Room Left'—Is That True?*, NBC NEWS (Oct. 29, 2019), https://www.nbcnews.com/better/lifestyle/travel-website-you-re-using-says-there-s-only-1-ncna1073066 [https://perma.cc/M7PU-MWBA] (reporting findings of consumer research group that travel websites create a false sense of scarcity by reporting availability of "obscure room types with low inventory"); *see also* Christopher S. Tang, *'Only 2 Left in Stock! Order Now!' But Does That Really Work?*, WALL ST. J. (Mar. 27, 2020), https://www.wsj.com/articles/only-2-left-in-stock-order-now-but-does-that-really-work-11585339621 [https://perma.cc/AL7E-A4A8] (explaining how scarcity messages can drive sales).

In yet another variation a site will pressure a consumer by offering "free two-day shipping" on the product "if you order in the next hour," failing to mention that two-day shipping is always available at no cost.[80] Endless variations are possible, but the theme is the same: create a false sense of urgency to pressure customers into purchases. The offer becomes all the more compelling when it is tailored to the exact item the customer wants to purchase, at exactly the time she is shopping for it.

Other techniques instead target consumers using fake news stories, endorsements from their favorite celebrities or politicians,[81] or fake posts from friends and social media contacts.[82] By tying their products to a consumer's social network, known news sources, or media personalities, scammers can overcome a consumer's ordinary caution toward unknown products and increase the likelihood that a scam will succeed—particularly when historical data is available to gauge which fake endorsements she is most likely to trust.[83] Examples of this type of scheme include a skin-care cream with a fake endorsements from Chelsea Clinton, Enhance Mind IQ Pills advertised alongside a fake story suggesting Elon Musk had appeared on *60 Minutes* to promote them, and an ad for colon cleansing supposedly endorsed by Kim Kardashian.[84]

---

[80] Amazon takes a similar approach in advertising its Amazon Prime service. At checkout, users are presented with a pop-up ad that encourages them to initiate an Amazon Prime subscription to obtain free shipping on the purchase. The ad makes no mention of Amazon's long-standing offer of free shipping for all orders over $25.

[81] Political alignment is a powerful predictor of consumer behavior. *See* Jihye Jung & Vikas Mittal, *Political Identity and the Consumer Journey: A Research Review*, 96 J. RETAILING 55, 55 (2020) (discussing political identity's important role in marketing); Eugene Y. Chan & Jasmina Ilicic, *Political Ideology and Brand Attachment*, 36 INT'L J. RSCH. MKTG. 630, 630 (2019) (explaining that "the growing division between political liberals and conservatives [may] mean[] a division in . . . consumption choices"); Jeremy B. Merrill, *Which Way Do You Vote? Facebook Has an Idea: Categorizing Users' Political Leanings*, N.Y. TIMES, Aug. 24, 2016, at A13 (discussing how Facebook determines users' political views the value of that information to advertisers).

[82] *Scams Starting on Social Media Proliferate in Early 2020*, FED. TRADE COMM'N: DATA SPOTLIGHT BLOG 2 (Oct. 2020), https://www.ftc.gov/system/files/attachments/blog_posts/Scams%20 starting%20on%20social%20media%20proliferate%20in%20early%202020%20/data_spotlight_oct_ 2020.pdf [https://perma.cc/Q3US-5A55] (observing that by "pretend[ing] to be someone you know" scammers "can get into a virtual community you trust"); Zeke Faux, *How Facebook Helps Shady Advertisers Pollute the Internet*, BLOOMBERG, https://www.bloomberg.com/news/features/2018-03-27/ad-scammers-need-suckers-and-facebook-helps-find-them [https://perma.cc/L6LT-TUZW] (Mar. 28, 2018) (interviewing top scam artists who report that despite their simplicity, "[f]ake personal endorsements and news reports are still the most effective tricks").

[83] Faux, *supra* note 82 (explaining that although scammers once had to guess which fake endorsements would be most effective with different demographic groups, "[n]ow Facebook does that work for them" by "track[ing] who clicks on the ad and who buys the pills" and then "target[s] others whom its algorithm thinks are likely to buy").

[84] *Id.*; *cf.* Compl. at 14–17, Fed. Trade Comm'n v. Effen Ads, LLC, No. 2:19-cv-00945-RJS (D. Utah Nov. 26, 2019), ECF No. 2 (describing the online promotion of fraudulent work-from-home

3. Vulnerable Populations

Still, other schemes use consumer data to target members of vulnerable populations. Service members, for example, are common targets for online scams because they move frequently, meaning they often sign up for services with new, unknown providers; they earn reliable pay and have access to GI Bill educational benefits;[85] and often they are young, with little experience managing finances.[86] Once browsing history or location data reveal a user to be a member of the armed-services community, she can be targeted for military-specific scams.

The bogus military-recruiting websites operated by Sunkey Publishing are a good example.[87] To target web users interested in military careers, Sunkey relied on Google's advertising tools to place ads for its websites alongside the search results of users whose searches suggested an interest in the military.[88] Users who clicked those ads were then directed to one of Sunkey's fake recruiting websites, like the one below, at domains including armyenlist.com, navyenlist.com, and airforceenlist.com.[89]

---

scheme using fake endorsements by the likes of Donald Trump and Warren Buffet and fake news stories purportedly by Fox News, CNN, and USA Today, among others).

[85] *See* Post-9/11 Veterans Educational Assistance Act of 2008, Pub. L. No. 110-252, 122 Stat. 2357 (2008) (codified at 33 U.S.C. §§ 3301–3324) (providing veterans with "enhanced educational assistance benefits").

[86] *See Veterans Consumer Protection: Preventing Financial Exploitation of Veterans and Their Benefits: Hearing Before the S. Comm. on Veterans' Affairs*, 118th Cong. 54–70 (2023) (statement of Monica Vaca, Deputy Director of the Bureau of Consumer Prot., Fed. Trade Comm'n).

[87] Compl. at 7–8, United States v. Sunkey Publ'g, Inc., No. 3:18-cv-01444-HNJ (N.D. Ala. Sept. 6, 2018), ECF No. 1 [hereinafter *Sunkey*, Compl.]. The case was resolved by stipulated order imposing a civil penalty on Sunkey and permanently enjoining it from "misrepresenting, expressly or by implication" that it is "affiliated with, or endorsed by . . . any . . . branch or agency of the United States federal government." Stipulated Final Order for Permanent Injunction and Civil Penalty Judgment at 8, United States v. Sunkey Publ'g, Inc., No. 3:18-cv-01444-HNJ (N.D. Ala. Sept. 12, 2018), ECF No. 6. All descriptions of any defendants' activities are as alleged in the relevant complaint. *Id.*

[88] *Sunkey*, Compl., *supra* note 87, at 8–9.

[89] *Id.* at 7–9.

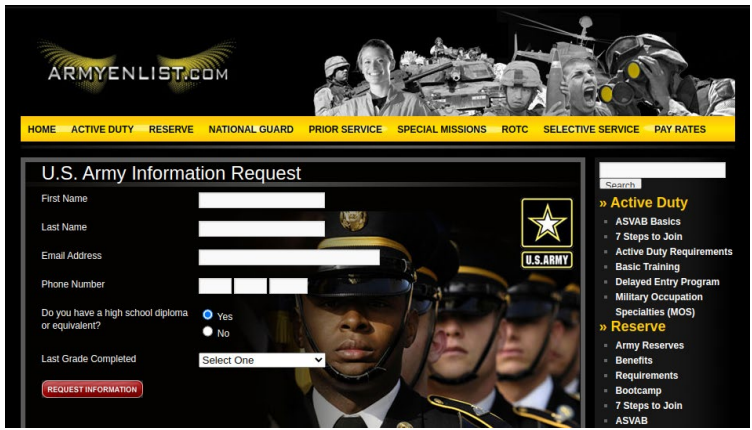

Figure 1. Main page for Sunkey's armyenlist.com website.[90]

Sunkey designed the pages to look like official recruitment websites of the U.S. military.[91] The armyenlist.com website, for example, included the U.S. Army logo and photos of military personnel and invited users to submit a "U.S. Army Information Request" by entering their name, contact information, and educational history into the provided web form.[92]

But Sunkey was not affiliated with the U.S. military or acting on behalf of any of its branches. Contact information submitted to the web form was directed to Sunkey's call center.[93] There, posing as official military recruiters, Sunkey's representatives telephoned the would-be recruits to gauge their interest not in military service, but in attending "military friendly colleges."[94] The military enlistment pages were just part of its larger effort to extract college-age visitors' contact information. Sunkey's real business was college referrals. Once candidates expressed interest in learning more about college opportunities, Sunkey transferred their information to colleges in exchange for lead-referral fees ranging from $15 to $40 per candidate.[95]

Other vulnerable groups, including children, retirees, and the disabled are also common targets. Sometimes a single scheme can even be adjusted to target multiple groups simultaneously. For example, from 2013 to 2018, MOBE Ltd. operated a fraudulent online business-education program, which purported to offer its customers a twenty-one-step program that would teach them to start and operate their own online businesses.[96] In reality, MOBE was a multilevel marketing scheme in which the vast majority of participants lost money.[97] The supposed educational programming consisted of sales pitches urging MOBE's customers to pay additional money into the scheme to earn higher commissions on referrals.[98]

---

[90] *Army Enlist*, SUNKEY PUBL'G, INC., http://www.armyenlist.com [https://web.archive.org/web/20120108174256/http://www.armyenlist.com/] [hereinafter *Army Enlist*].

[91] *Sunkey*, Compl., *supra* note 87, at 10–12.

[92] *Army Enlist*, *supra* note 90.

[93] *Sunkey*, Compl., *supra* note 87, at 31.

[94] *Id.* at 32–33.

[95] *Id.* at 34.

To entice customers, MOBE used specialized banner ads designed to target particular groups. For members of the military, it advertised the program as the "Patriot Funnel System" with an ad depicting a healthy, mid-career couple in front of an American-flag-themed backdrop.[99]

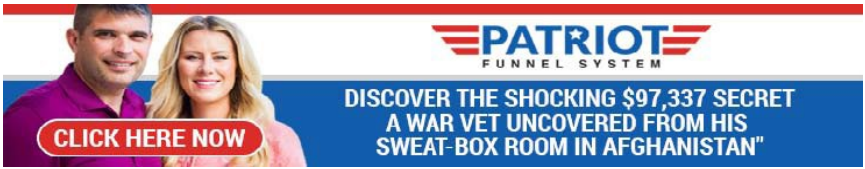

Figure 2. MOBE Ltd. banner ad targeting military members.

Clicking on the banner directed customers to the website, patriotfunnelsystem.com, which MOBE created specifically to target those in the military community.[100] There, MOBE invited visitors to provide their contact information for access to its "Top Secret FREE Video" and to "Discover How A War Veteran Uncovered The Secret to Earning Up To $3,300/day From His Sweat-box Living Quarters In Afghanistan."[101] The site told visitors they could, "Work when & where you want," and that, "It's as easy as copy and paste! No experience required!"[102] Following the links invited visitors into the same multilevel marketing scheme as everyone else.

Ads targeting other groups were tailored differently. For retirees, the scheme was pitched as the "Ultimate Retirement Breakthrough," a "Surefire Way To Create A Six-Figure Retirement Income In Less Than 12 Months."[103]

---

[96] Compl. at 2, Fed. Trade Comm'n v. MOBE Ltd., No. 6:18-cv-00862-RBD-DCI (M.D. Fla. June 4, 2018) [hereinafter *MOBE*, Compl.]. MOBE did not defend the action. *See* Fed. Trade Comm'n v. MOBE Ltd., No. 6:18-cv-862-Orl-37DCI, 2020 WL 3250220, at *7 (M.D. Fla. Mar. 26, 2020) (report and recommendation on motion for default judgment), *adopted*, 2020 WL 1847354 (M.D. Fla. Apr. 13, 2020) (adopting decision and permanently enjoining MOBE from marketing or selling its business-coaching programs).

[97] *MOBE*, Compl., *supra* note 96, at 19.

[98] *Id.* at 17–19.

[99] *Id.* at 24–25 fig.5.

[100] *Id.* at 23–24.

[101] *Patriot Funnel System*, MOBE LTD., https://patriotfunnelsystem.com [https://web.archive.org/web/20170308004811/https://patriotfunnelsystem.com/?aff_id=1760].

[102] *Id.*

[103] *MOBE*, Compl., *supra* note 96, at 25.

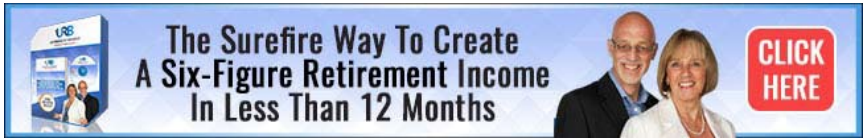

Figure 3. MOBE Ltd. banner ad targeting retirees.

Clicking the banner brought customers to another site, ultimateretirementbreakthrough.com, where they were shown pictures of carefree, retirement-age couples skiing, celebrating with grandchildren, and drinking wine.[104] The site promoted a "NO Savings Retirement Plan" to "Create Your Dream Retirement Lifestyle In 12 Months" with "3 Simple Steps," linked, of course, to the same MOBE program.[105]

MOBE even developed a version of the scheme to target those with physical and mental disabilities, touting the story of a "poisoned, brain-damaged man" who "rakes in a 6-figure income from home" and "Can Help You Earn Up To $1000 in Daily Commissions Working From Home As Little As 60 Minutes Per Day."[106] "If He Can Do It . . . You Can Too!", the site claimed, "It's as easy as copy and paste!"[107]

* * *

At one level, the new era of data-driven scams is just more in the long history of commercial deception that dates from time immemorial:[108] misrepresent yourself as affiliated with some trusted entity, gain the victim's trust, then reap the ill-gotten gains. History's con artists mastered this to perfection. Consider George C. Parker, who famously sold to unwitting buyers American landmarks, such as the Brooklyn Bridge, Grant's Tomb, and the Statue of Liberty, despite having no ownership interest in the properties.[109] It took a winning personality, the eye to spot a potential victim, and, importantly, an enor-mous amount of work. He learned to play the part of someone seemingly authorized to sell the properties (in the case of Grant's Tomb, the general's grandson); he set up a fake real estate office to run his swindles; and to complete the transactions, he prepared forged documents to evidence his ownership of the properties.[110] His careful work was rewarded with great success and, ultimately, life in prison.

---

[104] *Ultimate Retirement Breakthrough*, MOBE LTD., https://ultimateretirementbreakthrough.com [https://web.archive.org/web/20170603214756/https://ultimateretirementbreakthrough.com/?aff_id=1760].

[105] *Id.*; *MOBE*, Compl., *supra* note 96, at 25.

[106] *MOBE*, Compl., *supra* note 96, at 23–25; *Internet Funnel System*, MOBE LTD., https://internetfunnelsystem.com [https://web.archive.org/web/20160205042747/https://internetfunnelsystem.com/?aff_id=1760].

[107] *Internet Funnel System*, MOBE LTD., *supra* note 106.

[108] See *supra* Part I.A and accompanying notes, especially Dalley, *supra* note 41. For more on the common law's historical evolution in response to fraudulent market tactics, see generally Mark P. Gergen, *A Wrong Turn in the Law of Deceit*, 106 GEO. L.J. 555 (2018); Gregory Klass, *The Law of Deception: A Research Agenda*, 89 U. COLO. L. REV. 707 (2018); Page Keeton, *Fraud: The Necessity for an Intent to Deceive*, 5 UCLA L. REV. 583 (1958); W. Page Keeton, *Actionable Misrepresentation: Legal Fault as a Requirement* (pts. 1 &2), 1 OKLA. L. REV. 21 (1948), 2 OKLA. L. REV. 56 (1949); Leon Green, *Deceit*, 16 VA. L. REV. 749 (1930).

[109] *See* Gabriel Cohen, *Urban Tactics: For You, Half Price*, N.Y. TIMES, Nov. 27, 2005, at Q4 (telling the story of George C. Parker and other similar fraudulent sellers of famous landmarks).

Parker's strategy worked on the same, trusting human dispositions as modern internet scams, many of which are just as convincingly executed. Modern internet scams are no more persuasive than their physical-world counterparts. Kardashian-endorsed colon cleanser is just the newest snake oil. What makes modern internet schemes so different is the ease and accuracy with which scammers are now able to target their victims. Whereas it might once have taken Parker a lifetime to hone his quick wit and persuasive manner and a week at the dockyard to scout out prospective victims, most of the work can now be done by data and algorithm. There is no need to ask around for folks in the market for a new bridge; just check their browser search histories. Nor is there any need to hunt for gullible victims; the data will find those for you too. The result is a world full of the same old tricks, but in which those schemes can be run at much lower cost to would-be tricksters.

## II. Legislative and Regulatory Efforts

With 95% of Americans now online[111] and 90% owning a smartphone,[112] digital advertising has by now become commonplace. Indeed, given Americans' massive appetite for digital media[113] and their embrace of online shopping, digital advertising is the *premier* method for businesses to reach potential customers, especially for adults under fifty.[114] For consumers, that means more relevant ads and tailored brand experiences;[115] for sellers, it means better-targeted and cheaper advertising campaigns with higher customer realization rates.[116]

---

[110] *See The Daring Con Man Who Sold the Brooklyn Bridge*, EPHEMERAL N.Y. (Mar. 27, 2013), https://ephemeralnewyork.wordpress.com/2013/03/27/the-daring-con-man-who-sold-the-brooklyn-bridge/ [https://perma.cc/CBJ2-VDTK] (recounting Parker's cons and use of a fake real estate office).

[111] *Internet, Broadband Fact Sheet*, PEW RSCH. CTR., https://www.pewresearch.org/internet/fact-sheet/internet-broadband [https://perma.cc/U53S-LU8F] (Feb. 1, 2024) (documenting rise in percentage of U.S. adults who say they use the Internet from 52% in 2000 to 95% in 2023).

[112] *Mobile Fact Sheet*, PEW RSCH. CTR. (Jan. 31, 2024), https://www.pewresearch.org/internet/fact-sheet/mobile [https://perma.cc/NX6W-8EHZ] (showing poll results that as of 2023, 97% of Americans owned a cellphone of some kind and 90% owned a smartphone, up from 35% in 2011).

[113] *See* Andrew Perrin & Sara Atske, *About Three-in-Ten U.S. Adults Say They Are 'Almost Constantly' Online*, PEW RSCH. CTR., https://www.pewresearch.org/short-reads/2021/03/26/about-three-in-ten-u-s-adults-say-they-are-almost-constantly-online [https://perma.cc/PNP6-MYG6] (Mar. 26, 2021) (finding that, as of 2021, "31% of U.S. adults . . . go online 'almost constantly,'" which had increased from 21% in 2015).

[114] *See* Michelle Faverio & Monica Anderson, *For Shopping, Phones Are Common and Influencers Have Become a Factor—Especially for Young Adults*, PEW RSCH. CTR. (Nov. 21, 2022), https://www.pewresearch.org/short-reads/2022/11/21/for-shopping-phones-are-common-and-influencers-

But with the good comes also the bad. The widespread availability of cost-effective ad-targeting technology means it is also cheaper to run targeted scams. Thus, alongside traditional advertising, every manner of online deception has also flourished. Section A of this Part explains the technology-driven legislative approach that has been favored by lawmakers, and Section B discusses the drawback of that approach.

## A. Technology-Centric Responses to Deception

Legislatures,[117] regulators,[118] and law-enforcement officials[119] have all taken note of this problem. Yet thus far their efforts have met with only limited

---

have-become-a-factor-especially-for-young-adults [https://perma.cc/W9YD-39U7] (providing poll that shows 76% of American adults have made purchases on a smartphone and that 91% of American adults under fifty have done so). For a discussion of the Internet's evolution from its publication-centric roots in the 1990s to the complete virtual world and commercial center that it is today, see generally Gregory M. Dickinson, *Rebooting Internet Immunity*, 89 GEO. WASH. L. REV. 347, 354–72 (2021).

[115] Leslie K. John, Tami Kim & Kate Barasz, *Ads That Don't Overstep: How to Make Sure You Don't Take Personalization Too Far*, HARV. BUS. REV., Jan.–Feb. 2018, at 62 (explaining how "marketers have been able to gain unprecedented insight into consumers and serve up solutions tailored to their individual needs").

[116] *See* YAN LAU, FED. TRADE COMM'N, BUREAU OF ECONS., ECONOMIC ISSUES: A BRIEF PRIMER ON THE ECONOMICS OF TARGETED ADVERTISING 1, 6 (2020), https://www.ftc.gov/system/files/documents/reports/brief-primer-economics-targeted-advertising/economic_issues_paper_-_economics_of_targeted_advertising.pdf [https://perma.cc/6NEA-H4ZF] (explaining that precise targeting is a "crucial economic difference between internet advertising" and traditional advertising, and that "[p]recise targeting can lead to fewer ads being served [and] . . . lower marketing costs for firms by reducing wastage of ads served to disinterested customers").

[117] *See, e.g.*, Deceptive Experiences to Online Users Reduction (DETOUR) Act, H.R. 6083, 117th Cong. § 3(a) (2021) (prohibiting user interfaces "with the purpose or substantial effect" of undermining "user autonomy, decision making, or choice to obtain consent or user data"); California Consumer Privacy Act of 2018, CAL. CIV. CODE § 1798.140(h) (West 2024) (explaining that consent to data collection cannot be through "[a]cceptance of a general or broad terms of use" or obtained "through use of dark patterns"); Colorado Privacy Act, COLO. REV. STAT. § 6-1-1303(5)(a), (c) (2024) (same); Commission Regulation 2022/2065, of the European Parliament and of the Council of 19 October 2022 on a Single Market for Digital Services and Amending Directive 2000/31/EC (Digital Services Act), 2022 O.J. (L 277) 58 ("[O]nline platforms shall not design, organise or operate their online interfaces in a way that deceives or manipulates the recipients of their service.").

[118] *See, e.g.*, CAL. CODE REGS. tit. 11, § 7004(a) (2023) (implementing the California Consumer Privacy Act requiring, among other things, that designers of online interfaces "[a]void language . . . that [is] confusing to the consumer," "[a]void choice architecture that impairs or interferes with the consumer's ability to make a choice," or "add unnecessary burden or friction" to online processes); Trade Regulation Rule on Commercial Surveillance and Data Security, 87 Fed. Reg. 51273 (proposed Aug. 22, 2022) (requesting comments on companies' electronic privacy and data security practices, including tracking and targeted advertising); Negative Option Rule, 88 Fed. Reg. 85525 (proposed Dec. 8, 2023) (to be codified at 16 C.F.R. pt. 425) (providing notice of hearing on proposed amend-

success. The 40% increase in online shopping brought by the COVID-19 pandemic[120] was accompanied by a near 120% increase in reported instances of online shopping fraud.[121] This increase surely understates the true number of individuals who have fallen victim to online schemes. And the data show persistently elevated reports of online shopping fraud in subsequent years (459,627 in 2021; 365,549 in 2022; and 377,920 in 2023)[122] even as regulators have aggressively targeted online scammers.

---

ments to the FTC's Rule Concerning the Use of Prenotification Negative Option Plans); FED. TRADE COMM'N, BUREAU OF CONSUMER PROT., BRINGING DARK PATTERNS TO LIGHT 1–3 (2022), https://www.ftc.gov/system/files/ftc_gov/pdf/P214800%20Dark%20Patterns%20Report%209.14.2022%20-%20FINAL.pdf [https://perma.cc/3BQH-VRM6] (describing various types of dark patterns, explaining how they affect consumers, and discussing FTC efforts to combat them); Consumer Financial Protection Circular 2023–01: Unlawful Negative Option Marketing Practices, 88 Fed. Reg. 5727 (Jan. 30, 2023) (regarding restrictions on negative option marketing in the context of consumer financial products and services); Letter from Att'ys Gen. of Seventeen States & Dist. of Columbia, & Exec. Dir. of Hawaii, to the Fed. Trade Comm'n (Aug. 2, 2022), https://www.iowaattorneygeneral.gov/media/cms/17_Attorneys_General_Hawaii_OCP_Dig_FA07C81337A62.pdf [https://perma.cc/7M8H-MMJH] (documenting deceptive online commercial practices).

[119] *See, e.g.*, *Holiday Scams*, FED. BUREAU OF INVESTIGATION, https://www.fbi.gov/how-we-can-help-you/scams-and-safety/common-scams-and-crimes/holiday-scams [https://perma.cc/S5Q3-SE7U] (warning online consumers of potential fraud); *What to Do if You're Billed for Things You Never Got, or You Get Unordered Products*, FED. TRADE COMM'N: CONSUMER ADVICE (Aug. 2022), https://consumer.ftc.gov/articles/what-do-if-youre-billed-things-you-never-got-or-you-get-unordered-products [https://perma.cc/T3ZC-4SV9] (advising consumers on how to avoid online scams and how to respond); *Cryptocurrency Investment Scams*, U.S. SECRET SERV., CYBERCRIME INVESTIGATIONS, https://www.secretservice.gov/sites/default/files/reports/2022-11/cryptocurrency-investment-scams-v1.1.pdf [https://perma.cc/SY3W-C64D] (warning of common online scams related to cryptocurrency investment).

[120] *See* Mayumi Brewster, *Annual Retail Trade Survey Shows Impact of Online Shopping on Retail Sales During COVID-19 Pandemic*, U.S. CENSUS BUREAU (Apr. 27, 2022), https://www.census.gov/library/stories/2022/04/ecommerce-sales-surged-during-pandemic.html [https://perma.cc/89QA-VRNS] (reporting U.S. Census Bureau data that show a forty-three percent rise in online sales in the first year of the COVID-19 pandemic); John Koetsier, *E-Commerce Jumped 55% During Covid to Hit $1.7 Trillion*, FORBES, https://www.forbes.com/sites/johnkoetsier/2022/03/15/pandemic-digital-spend-17-trillion [https://perma.cc/73JX-7MX8] (Apr. 14, 2022) (reporting data from Adobe showing a rise of forty-one percent in 2020).

[121] Consumer complaints to the FTC of online fraud increased from 181,297 in 2019 to 397,735 in 2020. *See* FED. TRADE COMM'N, CONSUMER SENTINEL DATA BOOK 2022, at 84 (2023), https://www.ftc.gov/system/files/ftc_gov/pdf/CSN-Data-Book-2022.pdf [https://perma.cc/2JAG-L5MF] [hereinafter CONSUMER SENTINEL DATA BOOK 2022] (number of reports of online fraud made in 2020); CONSUMER SENTINEL DATA BOOK 2021, *supra* note 23 (number of reports of online fraud made in 2019).

[122] *See Fraud and ID Theft Maps*, *supra* note 23 (interactive report of fraud claims by time period and category); *see also* FED. TRADE COMM'N, CONSUMER SENTINEL DATA BOOK 2023, at 3 (2024), https://www.ftc.gov/system/files/ftc_gov/pdf/CSN-Annual-Data-Book-2023.pdf [https://perma.cc/YBL8-9ZY8] [hereinafter CONSUMER SENTINEL DATA BOOK 2023] ("Numbers change over time. The Sentinel Data Book sorts consumer reports by year, based on the date of the consumer's report. Some data contributors transfer their complaints to Sentinel after the end of the calendar year, and new data providers often contribute reports from prior years. As a result, the total number of reports

The proliferation of online scams can be attributed to three related factors.[123] First, centralized enforcement regimes like the FTC and its state analogues face an acute knowledge problem. Hundreds of thousands of instances of online trickery are reported each year, but many more surely go unreported. Try as they might, it is an enormous undertaking to police a large volume of widely dispersed instances of small-scale misconduct.

Second, it is surprisingly difficult to define with any precision what should count as a scam versus persuasive advertising or slick marketing. That difficulty limits the usefulness of two key tools in the regulatory arsenal: administrative rulemaking and official policy statements. Given the difficulty of precise proscription in this area, the FTC has traditionally eschewed formal rulemaking in favor of enforcement actions to police consumer deception.[124] But enforcement actions are expensive and time-consuming.

Third, and relatedly, the FTC and other enforcers have limited resources to pursue online wrongdoers. Indeed, the FTC initiated fewer than two-hundred consumer-protection actions in all of 2022, the vast majority of which were unrelated to online scams.[125] Given resource constraints,[126] regulators tend to devote what resources they have to only the highest profile and most egregious violators.

Pressed to act by the scale of online deception and yet facing impediments to their traditional tools, regulators have taken an alternative route—one which this Article refers to as a "tech-first" response. Identifying and pursuing online scammers is expensive and time-consuming. And no set of rules could

---

for 2023 will likely change during the next few months, and totals from previous years may differ from prior Consumer Sentinel Network Data Books. The most up-to-date information can be found online at ftc.gov/exploredata."). *Compare* CONSUMER SENTINEL DATA BOOK 2023, *supra* at 85 (providing the number of online-fraud complaints made each year over the past three years), *and* CONSUMER SENTINEL DATA BOOK 2022, *supra* note 121 (same), *with* CONSUMER SENTINEL DATA BOOK 2021, *supra* note 121 (same).

[123] *See generally* Gregory M. Dickinson, *Privately Policing Dark Patterns*, 57 GA. L. REV. 1633, 1649–61 (2023) (discussing the challenges faced by legislators and regulators attempting to combat dark patterns).

[124] *See* Maureen K. Ohlhausen, *The Procrustean Problem with Prescriptive Regulation*, 23 COMMLAW CONSPECTUS 1, 1–2 (2014) (FTC Commissioner Ohlhausen's explanation for the FTC's tradition of ex post enforcement as opposed to ex ante rulemaking).

[125] *See Legal Library: Cases and Proceedings*, FED. TRADE COMM'N, https://www.ftc.gov/legal-library/browse/cases-proceedings [https://perma.cc/D6C7-C34K] (supplying a database of FTC actions).

[126] *See* CONGRESSIONAL BUDGET JUSTIFICATION FISCAL YEAR 2024, *supra* note 23, at 8–9 (requesting a budget 2024 budget of $590 million, a 37% increase over the prior year, and 1,690 full-time positions); Lina M. Khan, Chair, Fed. Trade Comm'n, Testimony Before the House Appropriations Subcommittee on Financial Services and General Government 2 (May 18, 2022) (observing that the FTC's "total headcount today remains about two-thirds of what it was at the beginning of 1980").

ever distinguish lawful persuasion from unlawful coercion.[127] Indeed, that is precisely why the FTC was designed to regulate via case-by-case adjudication rather than administrative rulemaking. Defending the FTC Act in 1914 against the charge that its general language constituted an impermissible delegation of legislative authority, its principal sponsor, Senator Francis Newlands, explained that it would be a "hopeless task" to "define all the [dishonest] trade practices that can be devised," for they would simply "appear in some other form to-morrow."[128]

In contrast with the impossible task of defining "deception" ex ante, it is eminently possible for regulators to identify the tools that scammers use to perpetuate their deceptions—technologies like individualized consumer profiles, ad targeting, A/B testing, and dark-pattern-laden app and website interfaces. Not only is it easier to identify technologies than to define deception, it requires fewer resources. Instead of case-by-case adjudication, regulators and legislators can rely on prophylactic rulemaking to restrict all uses of problematic techniques. This approach would avoid the more labor-intensive process of case-by-case adjudication, which requires a showing in each instance that the technology, as deployed, has confused at least some substantial minority of reasonable consumers into entering an unwanted transaction.

Pressed by an avalanche of online deception and limited resources to combat it,[129] lawmakers have taken a tech-first approach that focuses on the technologies underlying scams rather than acts of deception themselves. For example, earlier this year Senators Josh Hawley and Richard Blumenthal introduced the No Section 230 Immunity for AI Act,[130] which would revoke online-intermediary immunity under 47 U.S.C. § 230[131] for claims predicated on an online entity's use of generative AI tools. By selectively eliminating a powerful defense, the bill would deter the use and development of all tools in that class of technologies, like ChatGPT and DALL-E,[132] that are "capable of generating novel text, video, images, audio, and other media."[133] The bill's authors are rightly concerned about the power of deepfakes and other sophisticated techniques that AI is now adding to mainstream scammers' toolboxes.[134] By regulating the tools of generative AI themselves, rather than "deepfakery" in particular, Senators Hawley and Blumenthal's bill takes the tech-first approach to solving that problem.[135]

---

[127] The FTC's current standard asks whether there has been a seller "representation, omission or practice that is likely to mislead the consumer acting reasonably in the circumstances, to the consumer's detriment." Letter from James C. Miller III, *supra* note 47.

[128] 51 CONG. REC. S11084 (daily ed. June 25, 1914) (statement of Sen. Francis G. Newlands). Rather than attempt the impossible, Newlands commended the example of Germany, which "leaves their tribunals to determine what practices are against good morals." *Id.*

[129] *See* Trade Regulation Rule on Commercial Surveillance and Data Security, 87 Fed. Reg. 51273, 51278–81 (proposed Aug. 22, 2022) (proposing that rulemaking recounting efforts to combat online deception through case-by-case enforcement and suggesting rulemaking given that the FTC's "limited resources today can make it challenging to investigate and act" on a case-by-case basis).

[130] No Section 230 Immunity for AI Act, S. 1993, 118th Cong. § 1 (2023).

[131] Communications Decency Act of 1996, Pub. L. No. 104-104, § 509, 110 Stat. 56 (1996) (codified at 47 U.S.C. § 230). For discussion of 47 U.S.C. § 230's history and ongoing controversies, see generally Dickinson, *supra* note 114.

Another example of Congress's tech-first approach to combating online deception is the DETOUR Act,[136] a bill introduced in 2021 by Senator Mark Warner and Representative Lisa Blunt aimed at reducing large online companies' ability to exploit online consumers' cognitive vulnerabilities. Among other things, the bill would bar companies from using A/B testing to hone their user interfaces by prohibiting companies from "segment[ing] consumers . . . into groups for the purposes of behavioral . . . research."[137] The concern is that companies will misuse A/B testing to develop deceptive interfaces that trick consumers.[138] Like the No Section 230 Immunity for AI Act, the approach targets a class of technology—A/B testing—rather than deceptive interfaces in particular, thus taking the easier tech-first approach.

Many state-legislative efforts have taken a similar path, particularly in response to "dark patterns"—website and smartphone user-interface designs like prechecked boxes or guilt-inducing language like, "No thanks, I don't like saving money," that can be used to press users toward particular choices. For ex-ample, the California Consumer Privacy Act[139] was amended in 2020[140] to prohibit companies from using "dark patterns" when collecting consumer consent to the use of personal information.[141] Similarly, the California Age-Appropriate Design Code Act,[142] enacted in 2023, imposes various design requirements on apps and websites "likely to be accessed by children" and expressly prohibits "[u]se of dark patterns to lead or encourage children to provide personal information beyond what is reasonably expected."[143] Dark patterns laws face significant challenges on First Amendment grounds as impermissible restrictions on commercial speech,[144] but what is important for present purposes is their tech-first structure: they target deception not directly, but by a broad prohibition on all interfaces deemed to be "dark patterns."

---

[132] *See* Rebecca Heilweil, *What Is Generative AI, and Why Is It Suddenly Everywhere?*, VOX (Jan. 5, 2023), https://www.vox.com/recode/2023/1/5/23539055/generative-ai-chatgpt-stable-diffusion-lensa-dall-e [https://perma.cc/D9AT-56HS] (providing a background on popular generative AI platforms).

[133] No Section 230 Immunity for AI Act, S. 1993 § 1(2).

[134] *See* Press Release, Sen. Josh Hawley, Hawley, Blumenthal Introduce Bipartisan Legislation to Protect Consumers and Deny AI Companies Section 230 Immunity (June 14, 2023), https://www.hawley.senate.gov/hawley-blumenthal-introduce-bipartisan-legislation-protect-consumers-and-deny-ai-companies-section/ [https://perma.cc/89QV-5FV7] (including deepfakes as part of the rationale for the bill).

[135] *See* Katie Paul, *Bipartisan U.S. Bill Would End Section 230 Immunity for Generative AI*, REUTERS, https://www.reuters.com/technology/bipartisan-us-bill-would-end-section-230-immunity-generative-ai-2023-06-14 [https://perma.cc/9QCJ-GSJ4] (June 14, 2023) (describing the bill's approach of regulating "emerging generative AI technology").

[136] Deceptive Experiences to Online Users Reduction (DETOUR) Act, H.R. 6083, 117th Cong. § 3(a) (2021).

[137] *Id.* § 3(a)(2).

[138] *See* Press Release, Sen. Mark W. Warner, Warner, Fischer Lead Bipartisan Reintroduction of Legislation to Ban Manipulative 'Dark Patterns' (July 28, 2023), https://www.warner.senate.gov/public/index.cfm/2023/7/warner-fischer-lead-bipartisan-reintroduction-of-legislation-to-ban-manipulative-dark-patterns [https://perma.cc/6W6A-KAB9] (explaining the DETOUR Act's purpose of preventing the use of dark patterns to "trick consumers").

### B. The Law-Tech Mismatch

Regulators' focus on new technologies is entirely predictable. It is only human, after all, to fear and to praise what is new more than what is commonplace and understood. That instinct is not far from the mark. Digital technologies have enabled persuasive advertising (and scams) at a scale not previously possible. New techniques may be no more deceptive than the timeworn tricks of skilled salespersons and con artists of old, but they certainly are cheaper to deploy.[145] Focusing on particular technologies and modes of deception has a superficial economic appeal. To regulators struggling under static budgets and rising online fraud, rulemaking to target problem-causing technologies may seem like the only option.

This shortcut, however, comes with major costs and takes the focus away from the less technical, more fundamental difficulties of digital deception: assimilation of dispersed knowledge, marshalling of enforcement resources, and procedural obstacles to collective action and private redress. This Section discusses the dangers of regulators' current, technology-focused response to digital deception through the examples of dark patterns, A/B testing, and genera-tive AI, setting the stage for discussion in the next Section of a series of legal obstacles that have hindered a robust private-law response.

---

[139] California Consumer Privacy Act of 2018, CAL. CIV. CODE §§ 1798.100–.199.100 (West 2024).
[140] California Privacy Rights Act of 2020, Proposition 24, 2020 Reg. Sess. (Cal. 2020) (codified as amended at CAL. CIV. CODE §§ 1798.100–.199.100 (West 2024)) (presented as Prop. 24, and approved by voters in the general election of Nov. 3, 2020).
[141] CAL. CIV. § 1798.140(h) (providing that "agreement obtained through use of dark patterns does not constitute consent").
[142] *Id.* §§ 1798.99.28–.40.
[143] CAL. CIV. § 1798.99.31(a)–(b) (effective Jan. 1, 2025).
[144] *See, e.g.*, NetChoice, LLC v. Bonta, 692 F. Supp. 3d 924, 966 (N.D. Cal. 2023) (preliminarily enjoining enforcement of the California Age-Appropriate Design Code Act), *aff'd in part, vacated in part*, 113 F.4th 1101 (9th Cir. 2024).
[145] *See supra* Part I.B–C.

1. Rules and Standards in Technology Regulation

Technology-focused regulation sounds sensible enough. If particular classes or applications of technology are helping online scammers target victims, why not subject those technologies to special restrictions? Sure, it is the scammers and not their tools who are blameworthy, but the goal is to stop online fraud, not to assign moral culpability. Either approach—catching scammers or taking their tools—will mean fewer victimized consumers; the tools are just easier to target.

Yet technology-specific restrictions come with significant drawbacks attributable to their tendency toward over- and under-inclusiveness. The problem relates to the well-known tradeoffs between legal rules and legal standards.[146] Like other legal rules, technology-based restrictions employ a highly definite test of applicability.[147] Determining whether a restriction applies requires resolution of only simple questions of fact; for example, whether the defendant used the specified technology. If so, she is subject to the restriction, otherwise, she is not. Definite rules of this sort are often preferable to flexible standards, which require qualitative human judgment,[148] because they make it possible for regulated parties to organize their affairs with greater certainty and at lower cost. For example, most drivers prefer posted speed limits to a prohibition on driving at an "unreasonable rate of speed."[149] Rules are preferable to standards whenever certainty is more important than accuracy, whereas standards, being adaptable, outperform rules in contexts that demand accuracy.

As discussed previously, however, legislators and regulators have tended toward technology-based rules in combating online deception rather than the standards-based ex post enforcement approach that the FTC has historically preferred.[150] Rather than a standard prohibiting entities from commercial methods likely to deceive reasonable app users into entering unwanted transactions, for example, proposed restrictions target various tools that might be employed to achieve that deception, including A/B testing, dark patterns, and generative AI.

---

[146] For detailed discussion of this concept, see Louis Kaplow, *Rules Versus Standards: An Economic Analysis*, 42 DUKE L.J. 557 (1992); Pierre Schlag, *Rules and Standards*, 33 UCLA L. REV. 379 (1985); Gideon Parchomovsky & Alex Stein, *Catalogs*, 115 COLUM. L. REV. 165, 172–81 (2015).

[147] *See generally* HENRY M. HART JR. & ALBERT M. SACKS, THE LEGAL PROCESS: BASIC PROBLEMS IN THE MAKING AND APPLICATION OF LAW 138–39 (William N. Eskridge, Jr. & Philip P. Frickey eds., Found. Press, Inc. 1994) (1958).

[148] *Id.* at 140 (defining standards as legal directives that require for their application, beyond findings of fact, "a qualitative appraisal of those happenings in terms of their probable consequences, moral justification, or other aspect of general human experience").

[149] *See* THOMAS A. LAMBERT, HOW TO REGULATE: A GUIDE FOR POLICYMAKERS 101 (2017) (discussing Montana's experiment with a common-law reasonableness standard for regulating speeding from 1995 to 1998).

[150] *See supra* Part II.A.

It would be one thing if the decision to rely on rules instead of standards were the result of reasoned deliberation concluding that certainty is more important than accuracy in this context. Unfortunately, the move toward technology-focused restrictions appears to be driven not by any careful deliberation, but by budgetary constraints, which favor easy-to-police rules over case-by-case evaluation of standards, and by the general pressure on regulators, when faced with a crisis, to do *something* to address the situation, even if available courses of action are suboptimal or even harmful. Indeed, in the context of online deception, across-the-board restrictions on particular techniques and technologies provide few of the benefits and all of the detriments of bright-line rules. They suffer from both over- and under-inclusiveness, sacrificing accuracy, while providing regulated entities with little corresponding certainty.

First, these restrictions foster uncertainty compared to rules' usual benefit of certainty. Under a rule-based regime, regulated parties can organize their affairs with confidence, perhaps investing heavily in a new product or business direction with certainty on which side of the law they stand. For example, automobile manufacturers have historically favored precise rules (that passengers are adequately protected if a vehicle is equipped with airbags or shoulder restraints) over the common-law requirement that products be "reasonably safe."[151] Both rules and standards aim in the same direction—fewer automobile injuries—but clear safety rules can be evaluated ex ante. This allows manufacturers to be certain they are on the correct side of the law early on, during development of their engineering and manufacturing strategies, rather than after the fact when those and other costs have been sunk.[152]

Much of the certainty provided by clear rules, however, is dependent on the rest of the legal regime in which they are embedded. A rule will always provide some degree more certainty than an equivalent standard, but even the clearest of rules cannot guarantee certainty where an entity is subject to other sources of legal uncertainty affecting the same choice.[153] For example, alt-hough Department of Transportation (DOT) regulations governing seatbelts and airbags gave manufacturers certainty regarding their federal obligations vis-à-vis the DOT, manufacturers argued vigorously in *Geier v. America Honda Motor Co.*[154] and *Williamson v. Mazda Motor of America, Inc.*[155] that those federal requirements should be interpreted not only to set federal standards but also to preempt contradictory or even supplementary state law.[156] Their concern was that the certainty accorded by the DOT's clear rules would be undermined if, notwithstanding compliance with federal regulations, they could face liability under the less predictable standards of state products liability law.[157] The certainty value of rules depends as much on their exclusivity as it does on their clarity.

---

[151] *See* RESTATEMENT (THIRD) OF TORTS: PRODS. LIAB. § 2(b) (AM. L. INST. 1998) (providing that "[a] product is defective in design when the foreseeable risks of harm posed by the product could have been reduced or avoided by the adoption of a reasonable alternative design . . . and the omission of the alternative design renders the product not reasonably safe").

[152] For a detailed discussion of how legal uncertainty can encourage inefficient overcompliance, especially when paired with large penalties for noncompliance, see Richard Craswell & John E. Calfee, *Deterrence and Uncertain Legal Standards*, 2 J.L. ECON. & ORG. 279, 292–95 (1986).

[153] *Cf.* Richard A. Epstein, *What Tort Theory Tells Us About Federal Preemption: The Tragic Saga of* Wyeth v. Levine, 65 N.Y.U. ANN. SURV. AM. L. 485, 490 (2010) (explaining how multi-

For online deception, governance exclusively by clear rules is impossible. Legislators and regulators could never hope to craft clear rules to define the entire universe of consumer deception or even the smaller range of abuses for a particular technology.[158] Restrictions on A/B testing or generative AI will inevitably be supplemented with the FTC Act's more general standard prohibiting unfair or deceptive acts or practices,[159] fifty states' similar (but not identical) Unfair and Deceptive Acts and Practices statutes,[160] and, of course, the eternally flexible common law of fraud, which allows recovery for any material misrepresentation that induces a consumer's justifiable reliance.[161] Clear rules re-

---

layered legal regimes can be counterproductive, for if one regime is overly restrictive, adding another set of restrictions can only make the problem worse); *see also* Steven Shavell, *A Model of the Optimal Use of Liability and Safety Regulation*, 15 RAND J. ECON. 271, 276–77 (1984) (identifying circumstances in which dual regulatory and tort regimes will be preferable to either alone).

[154] Geier v. Am. Honda Motor Co., 529 U.S. 861 (2000) (holding that Department of Transportation (DOT) airbag rule implicitly preempted common law action).

[155] Williamson v. Mazda Motor of Am., Inc., 562 U.S. 323, 336–37 (2011) (holding that DOT seatbelt rules did not preempt common-law action).

[156] The preemptive effect of the National Traffic and Motor Vehicle Safety Act of 1966, Pub. L. No. 89-563, 80 Stat. 718, (originally codified at 15 U.S.C. §§ 1381–1431 (1966)) (recodified without substantive change at 49 U.S.C. §§ 30101–30183 (1994)) and DOT regulations have been a topic of extensive controversy. The leading cases are *Geier*, 529 U.S. at 861, and *Williamson*, 562 U.S. at 323.

[157] Brief for Alliance of Automobile Manufacturers et al. as Amici Curiae Supporting Respondents at 22, Williamson v. Mazda Motor of Am., Inc., 562 U.S. 323 (2011) (No. 08-1314) (arguing that "[t]he need for clear and understandable rules of preemption cannot be overstated" given an automaker's need for clarity "before it produces a car," not years later after litigation).

[158] *See* Dickinson, *supra* note 123, at 1656–59.

[159] 15 U.S.C. § 45(a)(1).

[160] *See generally* Henry N. Butler & Joshua D. Wright, *Are State Consumer Protection Acts Really Little-FTC Acts?*, 63 FLA. L. REV. 163 (2011) (recounting the history of state Unfair or Deceptive Acts or Practices statutes and discussing differences in scope between state and federal law).

[161] RESTATEMENT (THIRD) OF TORTS: LIAB. FOR ECON. HARM § 9 (AM. L. INST. 2020) ("One who fraudulently makes a material misrepresentation of fact, opinion, intention, or law, for the purpose of inducing another to act or refrain from acting, is subject to liability for economic loss caused by the other's justifiable reliance on the misrepresentation.").

stricting particular techniques or technologies could thus communicate that businesses must not use those tools, or that they must not use them in a specified way. Rules cannot, however, give regulated entities ex ante certainty of compliance.[162]

Second, across-the-board restrictions for online deception are overinclusive. What rules give in clarity they take in accuracy. That is the tradeoff. Taking up the example offered earlier, a speed-limit statute that provides "thou shalt not exceed fifty-five miles per hour" offers certainty. But it comes at a cost. What of the professional driver on an empty highway in her well-maintained, low-center-of-gravity roadster, speedometer showing sixty? A common law reasonableness standard can accommodate her case, but not the speed-limit rule.

The same tradeoff occurs with technology. Almost all technologies have both desirable and undesirable uses. Uniform restrictions on all uses impede the good along with the bad and thus act as a tax (or prohibition) on desirable uses. Often that is a cost society chooses to bear. Prescription drugs are a good example. Drugs have tremendous capacity to treat illness and manage pain; they have been nothing short of a miracle for public health. But many drugs, especially narcotics, also have significant potential for abuse. To combat that risk, prescription narcotics are subject to across-the-board rules regulating their manufacture, distribution, and access—including for socially desirable uses.[163] Even though those regulations target narcotic prescription drugs as an entire class, there is little risk of overinclusiveness because the category of socially desirable uses of narcotics is so narrow. Put differently, there is significant overlap between the "technology" of narcotics and those applications of the technology that are desirable targets of regulation (the situation might be different were narcotics also an important fuel source or fertilizer, for example, but they obviously are not).[164] Their desirable uses being so limited, narcotics are well governed by bright-line rules applying to all uses.

---

[162] Rules can, however, give parties ex ante certainty of *noncompliance*, as with the FTC's negative-option marketing rule, discussed *infra* note 173. At the extreme are strict-liability rules, which provide certainty of liability for harms inflicted and may be preferable to reasonableness standards where information costs, litigation costs, or other transactions costs prevent sufficient enforcement to force actors to internalize the costs of their behaviors. For discussion of this approach in the context of the FTC's policing of data-security practices, see James C. Cooper & Bruce H. Kobayashi, *Unreasonable: A Strict Liability Solution to the FTC's Data Security Problem*, 28 MICH. TECH. L. REV. 257, 279–96 (2022).

[163] *See* Controlled Substances Act, Pub. L. No. 91-513, 84 Stat. 1242 (codified as amended at 21 U.S.C. §§ 801–904) (noting that many drugs "have a useful and legitimate medical purpose," but nevertheless creating broad restrictions on their use).

[164] *See generally* DANA A. SHEA, SCOTT D. SZYMENDERA & DAVID M. BEARDEN, CONG. RSCH. SERV., R43070, REGULATION OF FERTILIZERS: AMMONIUM NITRATE AND ANHYDROUS AMMONIA

There is no similar match between consumer deception and the technologies that support it. The technologies that support online deception are the same tools that drive any business endeavor: communications tools, computer databases, smartphones, and interactive user interfaces. Skilled entrepreneurs use them all to innovate, persuade, delight, and, sometimes, deceive their customers. Technological advancements in these tools, such as data mining, consumer profiling, A/B testing, and generative AI may seem like attractive targets for regulation, as was the printing press in centuries past,[165] but laws targeting the technologies themselves, rather than specific, undesirable uses, cannot help but be overinclusive. Beneficial uses will be impeded along with the bad.

Finally, just as a rule's accuracy can suffer from overinclusiveness, so too can it suffer from underinclusiveness. Federal drug laws are again a good example. In 1970, Congress enacted the Controlled Substances Act (CSA)[166] to combat the problems of drug abuse that had been rising throughout the 1960s.[167] The CSA identifies numerous problematic chemicals, from 3,4-methylenedioxy amphetamine to trimeperidine, which it classifies as controlled substances and divides into five schedules.[168] Schedule I substances are the most dangerous, having "high potential for abuse," "no currently accepted medical use," and no "accepted safety for use," and Schedule V the least, having "low potential for abuse," some "currently accepted medical use," and limited potential for dependence relative to other controlled substances.[169]

Federal law enforcement officials found early success enforcing the CSA in the 1970s and 1980s battling the Cocaine Cowboys and other distributors of controlled substances such as cocaine, marijuana, and heroin.[170] Their success, however, brought a new problem. As an alternative to more traditional drugs, suppliers began creating new synthetic analogues to those drugs that had the same psychoactive properties but were not listed as controlled substances un-

---

(2013) (highlighting that ammonium nitrate, while also a potentially dangerous substance, has an important usage as fertilizer and therefore is permitted under strict regulation).

[165] For a detailed history of free speech, the printing press, and their regulation, see generally JACOB MCHANGAMA, FREE SPEECH: A HISTORY FROM SOCRATES TO SOCIAL MEDIA (2022).

[166] Controlled Substances Act § 101.

[167] *See* H.R. REP. NO. 91-1444, pt. 1, at 1 (1970) ("This legislation is designed to deal in a comprehensive fashion with the growing menace of drug abuse in the United States."); Evelyn L.A. Jackson, Note, *Safe Injection Facilities: Reconsidering American Drug Policy*, 63 B.C. L. REV. 1467, 1479 (2022) (discussing federal criminalization of drugs starting in the 1960s, including passage of the Controlled Substances Act as part of the War on Drugs).

[168] 21 U.S.C. § 812(c).

[169] *Id.* § 812(b).

[170] *See History: 1980–1985*, DRUG ENF'T ADMIN. 49, https://www.dea.gov/sites/default/files/2021-04/1980-1985_p_49-58.pdf [https://perma.cc/3GV9-EJ62] (recounting a portion of the federal Drug Enforcement Administration's efforts to combat drug distribution).

der the CSA and thus lawful to manufacture, possess, and use.[171] A cat-and-mouse game began where the slow regulatory machinery of the FDA struggled, and failed, to ban new controlled substances as quickly as they could be created in chemical labs.[172]

To combat the new scourge of analogue drugs, Congress took an entirely different approach when it enacted the Controlled Substances Analogue Enforcement Act of 1986 (Analogue Act).[173] In contrast with the CSA's clear rules restricting individual chemicals by name, the Analogue Act introduced a flexible standard to regulate "controlled substance analogue[s]," which it defines as those substances, "the chemical structure of which" and the "stimulant, depressant, or hallucinogenic effects" of, are "*substantially similar to*" schedule I or II controlled substances.[174] It became necessary to adopt a standard alongside the CSA's existing scheduling rules because rules alone could not cover the universe of socially undesirable chemicals. In the drug context, rules are inherently underinclusive. It is simply too easy to design new drugs with slightly varied formulas that will skirt bright-line rules.

The same is true of consumer deception, where a similar cat-and-mouse game is forever at play. Product and service offerings, contracts, and the human language from which they are constructed are eminently flexible, far more so even than chemical structures. As with the CSA's list of controlled substances, targeted rules banning specific techniques of deception may sometimes be useful. The FTC's rules for negative-option marketing are a good example.[175] Negative-option marketing, where consumers are asked to opt out rather than opt in to transactions, is surely underinclusive of all consumer de-ception techniques; and rules against it will press scammers toward other methods. But, like controlled substances, the universe of socially desirable uses for negative-option marketing is limited, so not much harm is done by the rule, and it might reduce an especially noxious practice. The same cannot be said of broad rules limiting specific, multipurpose technologies. Such restrictions are simultaneously underinclusive of techniques for consumer deception, and therefore of limited use in stopping online scams, yet broad enough in scope to impede many socially desirable uses of those same technologies.

---

[171] *See* Gregory Kau, Comment, *Flashback to the Federal Analog Act of 1986: Mixing Rules and Standards in the Cauldron*, 156 U. PA. L. REV. 1077, 1086–87 (2008) (explaining the federal Controlled Substances Analogue Enforcement Act's (Analogue Act) standards-based approach to this problem); *see also* K. Michael Moore, *Fast Times in Federal Court and the Need for Flexibility*, 86 FORDHAM L. REV. 1681, 1681–82 (2018) (explaining the proliferation of synthetic designer drugs and the legislative response); Andrew Payne Norwood, Note, *Criminal Law—When Apples Taste Like Oranges, You Cannot Judge a Book by Its Cover: How to Fight Emerging Synthetic "Designer" Drugs of Abuse*, 39 U. ARK. LITTLE ROCK L. REV. 323, 324 (2017) (discussing the standards-based approach of the Analogue Act); Jeremy Mandell, Note, *Tripping Over Legal Highs: Why the Controlled Substances Analogue Enforcement Act Is Ineffective Against Designer Drugs*, 2017 U. ILL. L. REV. 1299, 1300–01 (explaining the designer-drug loophole, demand, and associated dangers).

[172] *See* Mandell, *supra* note 171, at 1307–08 (explaining that the regulatory process under the CSA is "far too slow to react to the speed and ingenuity of designer drug manufacturers" and that this "gave underground chemists a significant advantage").

[173] Controlled Substances Analogue Enforcement Act of 1986, Pub. L. No. 99-570, §§ 1201–1204, 100 Stat. 3207, 3207-13–3207-14 (1986) (amending various sections of the CSA, 21 U.S.C. §§ 801–904).

[174] 21 U.S.C. § 802(32)(A)(i)–(ii) (emphasis added).

[175] Negative Option Rule, 88 Fed. Reg. 85525 (proposed Dec. 8, 2023) (to be codified at 16 C.F.R. pt. 425).

2. Inaccuracy and Uncertainty in Action

The onslaught of cheap, targeted online deception has pressed regulators toward broad, across-the-board rules targeting the technologies that power scammers.[176] Such approaches are cheaper to implement than enhanced case-by-case enforcement against scammers. Yet, as the above analysis suggests, in this context, the choice between rules and standards has consequences far broader than enforcement costs.[177] Because of their over- and underinclusiveness, rules broadly restricting all uses of multipurpose technologies will impede all beneficial uses of those technologies for the minor benefit of pressing scammers toward alternative methods. And because regulated parties remain subject also to general prohibitions on unfair and deceptive practices under state and federal law, these costs will not be offset by rules' usual benefit of providing ex ante certainty of compliance. This argument is a general one, applying to any across-the-board rule restricting multipurpose technology,[178] but it is nonetheless helpful to consider a few specific examples.

First, consider the DETOUR Act's proposed restrictions targeting A/B testing. Under the Act, large online companies would be prohibited from "segment[ing] consumers of online services into groups" for research (a prerequisite to A/B testing) unless they first obtain the "express, affirmative" consent of each user.[179] Consumers would be able to consent only by affirmatively opting in, instead of by "a general contract or service agreement" like those that typically accompany installation of an app or registration for a web ser-vice.[180] Moreover, companies would be required to re-inform users every ninety days of ongoing testing, establish independent review boards registered with the FTC to oversee any such testing, and issue quarterly public notices of any studies undertaken to promote user engagement.[181]

---

[176] *See supra* Part I–II.A.

[177] *See* Ohlhausen, *supra* note 124, at 3–6 (recommending "regulatory humility" in the face of technological change, as early skepticism of new technology is not an accurate predictor of harm, and ex ante rules produce unanticipated harm and bar unforeseen opportunities).

[178] Consider also, for example, proposed laws targeting generative AI, such as the Generative AI Copyright Disclosure Act of 2024, H.R. 7913, 118th Cong. (2024); the California AI Transparency Act, S.B. 942 (Cal. 2024); and the No Section 230 Immunity for AI Act, S. 1993, 118th Cong. (2024), discussed *supra* Part II.A and in notes 130–134.

[179] Deceptive Experiences to Online Users Reduction (DETOUR) Act, H.R. 6083, 117th Cong. § 2(6)(A), 3(a)(2).

These requirements would, of course, have a major chilling effect on A/B testing. Companies will be hesitant to issue announcements to users and the public that they engage in scary-sounding "psychological experimentation"; their users will be hesitant to agree, likely to instead reject the request without much thought; and whatever A/B testing does proceed would be less effective because it would be based on smaller, less representative samples of users. The end result would be much less, and much less effective user interface testing. Indeed, that is the Act's objective: to reduce companies' use of user interface designs that influence user choices by undermining companies' ability to research which interfaces users will find most persuasive.[182]

But this technology-focused rule would have its costs. To combat those companies who abuse A/B testing to develop tools that deceive users, the DETOUR Act would impose costs and limit the effectiveness of all A/B testing, even the multitude of uses that are socially beneficial: Google testing new functionality in Gmail to determine whether users find them intuitive or irritating; Amazon testing new product-recommendation algorithms; even Duolingo studying which language-practice nudges and animations motivate users to meet their learning goals.

Limits on A/B testing would also undermine online services more generally, by impairing their ability to monetize their products. To product designers, an "effective" user interface is one that the consumer enjoys using and that is also profitable for its creator. For the many apps and services that are offered for free, profitability requires a steady stream of advertising revenue. A/B testing is a key tool for companies to determine where and how many ads to include in their products to make their operations profitable and support future product enhancements without annoying and driving away their customers. Thus, beyond degrading user-interface designs, restrictions on A/B testing would reduce companies' profitability and ability to fund product improvements. In short, consumers would see fewer products and suffer poorer product performance,[183] while scammers would persist—either violating the law, as scammers are wont to do, or perpetrating their schemes in some other way.

---

[180] *Id.* § 2(6)(B)(ii).
[181] *Id.* § 3(b).
[182] *See* Kaitlyn Tiffany, *Some Apps Use Design to Trick You into Sharing Data. A New Bill Would Make That Illegal.*, VOX (Apr. 10, 2019), https://www.vox.com/the-goods/2019/4/10/18304781/social-media-dark-pattern-design-bill-facebook-ftc [https://perma.cc/Q7PD-9VFU] (explaining the DETOUR Act's objectives and effect on A/B testing).

Second, consider regulations restricting dark patterns. In an effort to combat online deception, a growing number of state legislatures have prohibited dark patterns in user interface designs.[184] As I have discussed elsewhere, the concept of dark patterns is so vague that it is difficult for lawmakers to craft legislation targeting them. In an effort to be comprehensive, such laws tend to include catch-all provisions that amount to instructions that user interfaces must not be "*too* tricky."[185] These catch-all provisions mean the laws provide little certainty to regulated entities, who can see no guaranteed path of compliance. And the laws are unnecessary because the conduct they target is already barred by general consumer protection laws prohibiting unfair or deceptive practices of any sort.

Insofar as dark pattern laws are mere regurgitations of existing consumer protection laws, they are unnecessary at best. To the extent dark patterns laws purport to do something more—to identify a class of technology or advertising techniques that should be categorically prohibited regardless of whether they constitute traditional consumer deception—their effects will be even worse. Such restrictions are prone to suffer the same over- and under-inclusiveness problems as all across-the-board regulation of any multipurpose technology.

The term "dark patterns" is applied to a host of user-interface designs ranging from those used to perpetrate outright fraud[186]—where app makers and website operators peddle lies to separate customers from their money—to or-dinary marketing gimmicks,[187] addictive apps,[188] and even app designs that are not intentionally malicious but are poorly designed and difficult to use.[189] Indeed, observing the lack of "a singular concern or consistent definition" in the dark-pattern literature one group of scholars has suggested facetiously that dark patterns are those "user interface designs that researchers deem problematic."[190]

---

[183] Recent empirical literature shows a reduction in online product quality and number of new product offerings where privacy laws reduce online companies' revenues by restricting the use of consumer data for targeted advertising. *See, e.g.*, Garrett Johnson, Tesary Lin, James C. Cooper & Liang Zhong, COPPAcalypse? The YouTube Settlement's Impact on Kids Content (Mar. 14, 2024), https://papers.ssrn.com/sol3/papers.cfm?abstract_id=4430334 [https://perma.cc/75SS-MAYJ]; Tobias Kircher & Jens Foerderer, Does Privacy Undermine Content Provision and Consumption? Evidence from Educational YouTube Channels (Jan. 19, 2024), https://papers.ssrn.com/sol3/papers.cfm?abstract_id=4473538 [https://perma.cc/4SLU-3FQF]; *see also* Daniel J. Gilman & Liad Wagman, *The Law and Economics of Privacy*, 29 UCLA J.L. & TECH. 55, 94–102 (2024) (summarizing recent work studying the effects of consumer privacy legislation on firm revenue, market concentration, product quality, and innovation).

[184] *See supra* Part II.A.

[185] Dickinson, *supra* note 123, at 1656–61. Thus, for example, California privacy regulations bar interfaces that impose "unnecessary burden or friction on users" or have the effect of "substantially subverting or impairing user autonomy, decisionmaking, or choice." CAL. CODE REGS. tit. 11 § 7004(a)(5), (c) (2024).

[186] *See, e.g.*, Arunesh Mathur et al., *Dark Patterns at Scale: Findings from a Crawl of 11K Shopping Websites*, 3 PROC. ASS'N FOR COMPUTING MACH. ON HUM.-COMPUT. INTERACTION 1, 2 (2019) (describing how dark patterns "trick users into signing up for recurring subscriptions and making unwanted purchases, resulting in concrete financial loss").

Laws prohibiting "dark patterns" thus put businesses in a difficult spot. They gain little clarity, remaining subject to general laws barring consumer deception, but they are presented with a new, vague prohibition on using "dark pattern" interface designs. Might Duolingo's tenacious efforts to encourage language learning through reminders and exciting, graphic animations "stimulate[] . . . repetitive behavior" or "impair[] user . . . choice"?[191] What about Netflix's and YouTube's autoplay features that begin playing the next video automatically? Those features mean that consumers need not track down the remote control to continue watching. They also affect user choices and contribute to binge watching. Might such features be prohibited even if they do not deceive users because they undermine their autonomy? No one knows. The end result of broad prohibitions on "dark pattern" techniques will be greater caution in new interface design, leading to fewer intuitive, new, and interesting designs. That is a steep price to pay for a belt-and-suspenders law that applies primarily to conduct already prohibited by existing laws.

---

[187] *See* Alex Bitter, *The Sneaky Ways Instacart Gets You to Buy More Stuff*, BUS. INSIDER (Feb. 16, 2023), https://www.businessinsider.com/how-instacart-uses-dark-patterns-encourage-shoppers-buy-faster-more-2023-2 [https://perma.cc/DV3L-YEQA] (website message reading, "Items in your cart are selling fast! Check out soon before they're sold out"); European Commission Press Release IP/23/418, Consumer Protection: Manipulative Online Practices Found on 148 Out of 399 Online Shops Screened (Jan. 30, 2023) ("[V]isual design or choice of language" that "directed consumers toward certain choices" including "more expensive products or delivery options").

[188] *See, e.g.*, Gregory Day & Abbey Stemler, *Are Dark Patterns Anticompetitive?*, 72 ALA. L. REV. 1, 2–3 (2020) (recounting Snapchat's use of "streaks" to keep users addicted to the app); *see also Addiction by Design: Dark UX Patterns*, ANKIT SHERKE DESIGN https://www.ankitsherke.design/blog/addiction-by-design-dark-ux-patterns [http://web.archive.org/web/20220817080526/https://www.ankitsherke.design/blog/addiction-by-design-dark-ux-patterns] (explaining that dark pattern designs "make you do things that you didn't mean to, like buying or signing up for something, or spending more time on these apps than you intended").

[189] *See* Devin Coldewey, *Instagram Gets Worse with Dark Patterns Lifted from TikTok*, TECHCRUNCH (July 25, 2022), https://techcrunch.com/2022/07/25/instagram-gets-worse-with-dark-patterns-lifted-from-tiktok/ [https://perma.cc/9RR8-LE2K] (redesign of Instagram app user interface to change scrolling and audio control behavior); Catherine Zhu, *Dark Patterns—A New Frontier in Privacy Regulation*, REUTERS (July 29, 2021), https://www.reuters.com/legal/legalindustry/dark-patterns-new-frontier-privacy-regulation-2021-07-29/ [https://perma.cc/22F7-7Q5Z] (describing techniques like gamification and nudging that might not be intended to deceive, but nonetheless "erod[e] user agency and could be deemed to be dark patterns"); Justin Hurwitz, *Designing a Pattern, Darkly*, 22 N.C. J.L. & TECH. 57, 60 (2020) (attributing many dark patterns to the difficulty of user-interface design); *cf.* Nick Doty & Mohit Gupta, *Privacy Design Patterns and Anti-Patterns: Patterns Misapplied and Unintended Consequences* 2 (2013), https://cups.cs.cmu.edu/soups/2013/trustbusters2013/Privacy_Design_Patterns-Antipatterns_Doty.pdf [https://perma.cc/45Q5-ZP8Y] (reasoning that although "some anti-patterns might be the perverse case," many result from the thoughtless deployment of established techniques to new and inappropriate contexts).

[190] *See* Arunesh Mathur, Jonathan Mayer & Mihir Kshirsagar, *What Makes a Dark Pattern . . . Dark?: Design Attributes, Normative Considerations, and Measurement Methods*, *in* PROCEEDINGS OF THE 2021 CHI CONFERENCE ON HUMAN FACTORS IN COMPUTING SYSTEMS, Article No. 360, at 1, 1 (2021).

[191] Deceptive Experiences to Online Users Reduction (DETOUR) Act, H.R. 6083, 117th Cong. §§ 2(4), 3(a)(1).

### III. Targeting the Real Tools of Deception

Online deception is a real problem. Its growth has outpaced governmental enforcement resources;[192] yet, because the digital technologies that drive it are so flexible and widely used, across-the-board restrictions are bound to suffer from severe over- and under-inclusiveness.[193] The natural question, then, is what *can* be done to combat the rise in online deception. This Part proposes a coordinated public-private enforcement scheme that would reallocate government enforcement resources to combat those varieties of online fraud that have been most resistant to private litigation.

#### A. The Role of Private Enforcement

Most discussions of online fraud have focused on public law: identifying the technologies driving deception and contemplating fresh legislation and regulations to help the government combat the problem. Largely missing from the conversation, however, has been the potential role for private litigation—lawsuits brought against scammers by their online-fraud victims under the traditional doctrines of property, tort, restitution, and contract law or under state consumer-protection statutes. The absence of private enforcement actions from the conversation is striking, as its features would seem to make it especially well-suited to the task.

First, private enforcement produces decisional law created by courts' adjudication of individual disputes on a case-by-case basis, after the incident in question, and with all available facts. Consider the tort of negligence, which imposes liability for causing physical harm to another through one's failure to "exercise reasonable care under all the circumstances."[194] Unlike a legislature establishing a forward-looking rule, a court determining whether a defendant's conduct amounts to negligence need not imagine beforehand all conceivable human conduct and categorize it as either reasonable or unreasonable. Instead, looking backward and considering all of the circumstances, a court decides whether the defendant's acts were consistent with the behavior of a reasonably prudent person.[195] As time passes and more cases are decided, trends in decisions will emerge and form precedents that guide future decisions. These decisions are crafted ex post, meaning that lawmakers are never forced to define the full scope of the law ex ante, which would require substantial technical expertise even to attempt,[196] and is an impossible and counterproductive task for online deception.

---

[192] *See supra* Part I.
[193] *See supra* Part II.A.
[194] Restatement (Third) of Torts: Liab. for Physical & Emotional Harm § 3 (Am. L. Inst. 2024).

Second, a related benefit of private enforcement is its power to assimilate the dispersed knowledge of the public and the wisdom of judges around the country. The private parties who are deceived by online scams, after all, have the most comprehensive access to information about those scams and a strong incentive to pursue their claims. And, unlike the government, the private sector has virtually unlimited enforcement resources in the form of private attorneys willing to pursue claims against wrongdoers. For all the consternation it causes to law students and civil lawyers, there is a beautiful method to the common law's madness: Even when the government fails to discover the fraud or, as is likely, lacks sufficient resources to bring an enforcement action against the perpetrator, the individuals who have been scammed can themselves file suit thereby bringing the matter to the court's attention and obtaining redress without any need to rely on FTC enforcers. Moreover, because private enforcement under state common law or state consumer-protection statutes would be pursuant to general standards prohibiting deception, rather than rules restricting particular technologies, use of those tools could continue unabated.

---

[195] *See* Giacomo A.M. Ponzetto & Patricio A. Fernandez, *Case Law Versus Statute Law: An Evolutionary Comparison*, 37 J. LEGAL STUD. 379, 379 (2008) ("Case law develops gradually through the rulings of appellate judges who have heterogenous preferences but are partially bound by stare decisis. We show that its evolution converges toward more efficient and predictable legal rules. Since statutes do not share this evolutionary property, case law is the best system when the efficient rule is time invariant, even if the legislature is more democratically representative than individual judges are."). Of course, regulatory enforcement actions form decisional law too. Indeed, that is how the FTC has generally policed this area. That body of decisional law guides both regulated entities and future enforcement actions, but given the FTC's limited enforcement resources, it is less robust and evolves more slowly than might judicial decisional law. Regardless, the bigger point here is that decisional law responding to dark patterns is superior to attempts by legislation or regulation to ban specific techniques or technologies.

[196] On the technical capacity of courts to resolve claims of fraud and misrepresentation, which "are familiar concepts in Anglo-American jurisprudence" the adjudication of which "should involve neither conceptual nor practical difficulties," see Richard A. Posner, *The Federal Trade Commission*, 37 U. CHI. L. REV. 47, 66-68 (1969). *But see* Kevin Thomas Frazier, *An "F" in Judicial Education: Why Emerging Technologies and New Risks Demand Judicial Education Reform*, 50 OHIO N.U. L. REV. 27 (2023) (observing that judges' lack of familiarity with emerging technologies may hamper their ability to serve as evidentiary gatekeepers).

Private enforcement also has its drawbacks, of course. Precedent-based decision making sacrifices ex ante certainty for ex post flexibility and provides less certainty to regulated parties.[197] The approach shines, however, in contexts where ex ante rules are impossible or undesirable, and thus, there is little loss of certainty to regulated parties by delaying classification of behavior as lawful or unlawful until after all facts are in hand. Online deception is such a context. Commercial shenanigans are as old as humanity; the internet is just their most recent playground. Quickly evolving misconduct that resists across-the-board classification is perfect for resolution through precedent-driven decisional law. Judges, empowered with the broad mandates like those of tort, contract, and consumer-protection statutes to root out material misrepresentations, justifiable reliance, and fraudulent inducements, would have exactly the tools they need to provide redress for online fraud without the societal losses that would result from across-the-board restrictions on the technological tools employed.

### B. The Patterns of Digital Deception

Unfortunately, private enforcement efforts against online deception have met with only partial success. Online deception is triply unlawful under the FTC Act, state consumer-protection statutes, and common-law fraud, but various procedural challenges to private litigation—what this Article calls the Patterns of Digital Deception—sometimes allow online scammers to avoid the private lawsuits that could otherwise act as a check on online fraud. This section categorizes those procedural obstacles with two aims in mind: first, so that the FTC and state regulators can refocus their efforts on those areas where private enforcement is least adept; and second, in the longer term, to aid lawmakers in crafting legislation to reduce those obstacles.

1. Fly-by-nighters

First are the *fly-by-nighters*. Many entities that perpetrate online scams make efforts to avoid detection. They may be based in foreign jurisdictions,[198] create fake online identities,[199] or conceal their phone numbers and IP address-es[200] to make their true identities and whereabouts difficult to discern.[201] Such cross-border fraud is on the rise both in absolute and relative terms. Whereas cross-border fraud reports constituted less than 1% of all fraud reports in 1996, that number has risen to more than 10% in 2023.[202] In the last five years, the FTC has received fraud complaints connected with 231 different countries,[203] most prominently China, which is connected with 25% of all cross-border fraud reports.[204]

---

[197] *See supra* Part II.B.
[198] *See* LINA M. KHAN, REBECCA KELLY SLAUGHTER & ALVARO M. BEDOYA, FED. TRADE COMM'N, THE U.S. SAFE WEB ACT AND THE FTC'S FIGHT AGAINST CROSS-BORDER FRAUD 3 (2023), https://www.ftc.gov/system/files/ftc_gov/pdf/ftc_safe_web_congressional_report_oct_2023.pdf [https://perma.cc/QQD9-ZEGC] [hereinafter FTC CROSS-BORDER FRAUD REPORT] (discussing the FTC's efforts to combat cross-border fraud).
[199] *See, e.g.*, *What to Know About Romance Scams*, FED. TRADE COMM'N: CONSUMER ADVICE (Aug. 2022), https://consumer.ftc.gov/articles/what-know-about-romance-scams [https://perma.cc/4L7P-D5Z8] [hereinafter *FTC Romance Scams*] (reporting that romance scammers create fake profiles on social media sites).

Not only does that complicate law-enforcement efforts, but it also makes it exceedingly difficult for private litigants to obtain relief. First, they must identify the entity that is behind the website, app, or email that facilitated the fraud. That is often a difficult task when dealing even with a domestic entity, let alone a foreign-based one that has made efforts to conceal its activities. Second, assuming the foreign entity can be identified and brought before a court in the United States, a private plaintiff faces significant procedural obstacles to having her claim heard. A forum-selection clause in any agreement between the consumer and the defendant may require litigation in a foreign jurisdiction; if not, the defendant is likely to object that the court lacks personal jurisdiction[205] over it and that, even were jurisdiction appropriate, that the forum non conveniens doctrine[206] counsels against litigation in the United States.

Even were litigation to proceed, the discovery process could require overseas travel to depose the defendant's employees, plus language-translation services to process written discovery and permit deposition of non-English speakers. Finally, even were a private plaintiff to obtain judgment, she will have difficulty enforcing it if the defendant shields its assets in foreign jurisdictions. Given that the median consumer loss from cross-border fraud is between $141 and $511,[207] it is easy to see why victims decline to pursue private relief through the court system. The legal obstacles are too many and the sums at stake too small to justify the effort.

---

[200] *See Caller ID Spoofing*, FED. COMMC'NS COMM'N, https://www.fcc.gov/spoofing [https://perma.cc/N5E4-U6AL] (Mar. 7, 2022) (discussing caller ID spoofing to conceal a caller's true phone number).

[201] *See* FTC CROSS-BORDER FRAUD REPORT, *supra* note 198, at 6 & n.24 (observing that "when reporting fraud, consumers may not even know that a foreign entity is involved" and that "between 18% and 39% of fraud complaints received by the FTC between 2019 and 2022 could not be identified as either cross-border or non-cross border").

[202] *Id.* at 1.

[203] *Id.* at 6.

[204] *Id.* at 7 fig.3.

[205] *See* 4A CHARLES ALAN WRIGHT & ARTHUR R. MILLER, FEDERAL PRACTICE AND PROCEDURE § 1073 (4th ed.) (personal jurisdiction over internet-based defendants).

[206] *See* 14D *id.* § 3828 (forum non conveniens doctrine).

[207] FTC CROSS-BORDER FRAUD REPORT, *supra* note 198, at 5.

2. Nickel-and-dimers

Second are *nickel-and-dimers*. Some online scams, particularly online romance scams[208] and Nigerian-prince-type email scams,[209] are famous for swindling victims out of vast sums, often entire retirement-savings accounts or even the victims' home. But many other types of scams involve sums that are simply too small to support individual private enforcement.[210]

One recent example is the credit-repair service offered by Turbo Solutions. The company claimed that it could improve consumers' credit scores by "200 points in 90 days"[211] so that they could "gain access to mortgages . . . [and] car loans."[212] The company claimed that it could "Delete Negative Accounts" from and provide "Credit Boosters" to customers credit histories to "help get you that approval that you need."[213] Rather than provide any legitimate services, however, the company set out to increase customers' credit scores by unlawful and ineffective means such as filing false identity-theft reports and baseless credit disputes.[214] For these services, Turbo Solutions charged an up-front fee of fifteen hundred dollars.[215]

That is a significant sum, especially for consumers who have poor credit and are struggling to manage their finances. Yet it is a far cry from the amount that would be required to justify and support private litigation through contingency fees. A consumer tricked by such a scheme will simply swallow her loss rather than throw good money after bad. The problem is worse still for the even smaller sums at issue when a consumer is tricked by a hidden shipping charge or purchases an online service deceptively set to automatically renew. In the aggregate, such schemes impose large losses on society by allocating resources to undesired or unlawful products and services. To the individual consumer, however, they are small potatoes—certainly not worth bringing a lawsuit over.

---

[208] *See FTC Romance Scams*, *supra* note 199 (romance scammers make fake social media profiles); Emma Fletcher, *Reports of Romance Scams Hit Record Highs in 2021*, FED. TRADE COMM'N: DATA SPOTLIGHT BLOG (Feb. 10, 2022), https://www.ftc.gov/news-events/data-visualizations/data-spotlight/2022/02/reports-romance-scams-hit-record-highs-2021 [https://perma.cc/HTW2-BUUY] (reporting that the extent of losses to romance scams are "more than any other FTC fraud category").

[209] *See* Press Release, Fed. Trade Comm'n, Western Union Admits Anti-Money Laundering Violations and Settles Consumer Fraud Charges, Forfeits $586 Million in Settlement with FTC and Justice Department (Jan. 19, 2017), https://www.ftc.gov/news-events/news/press-releases/2017/01/western-union-admits-anti-money-laundering-violations-settles-consumer-fraud-charges-forfeits-586 [https://perma.cc/L3BQ-45U8] (reporting Western Union's complicity with scams involving sending money overseas); Megan Leonhardt, *'Nigerian Prince' Email Scams Still Rake in Over $700,000 a Year—Here's How to Protect Yourself*, CNBC (Apr. 18, 2019), https://www.cnbc.com/2019/04/18/nigerian-prince-scams-still-rake-in-over-700000-dollars-a-year.html [https://perma.cc/67EY-NFB7] (explaining the "Nigerian Prince" scam and discussing its prevalence).

[210] Legislatures might overcome such obstacles via statute providing for attorney's fees, statutory damages, or other incentives to encourage more enforcement. *See* Dickinson, *supra* note 123, at 1665–68 (discussing these strategies and legislative tools to mitigate risks of overenforcement).

[211] Compl. at 10, United States v. Turbo Sols. Inc., No. 4:22-mc-00369 (S.D. Tex. Mar. 1, 2022).

[212] *Id.* at 8.

[213] *Id.* at 8–9.

[214] *Id.* at 2, 12.

[215] *Id.* at 10.

3. User-Interface Shapeshifters

Third, are *user-interface shapeshifters*. Small-value claims are not a uniquely modern problem. Two procedural tools are available to claimants whose small losses might otherwise discourage litigation. First is the availability of attorney's fees in some contexts.[216] A victim who has suffered a small loss cannot expect an attorney to represent her on a contingency basis; the potential recovery is simply too small. Nor would she pay out of pocket for an attorney's services for a prospective gain far smaller than the loss. She might be willing to pursue recovery, however, where attorney's fees are available to successful litigants. That approach has proven successful, for instance, in the case of the Civil Rights Act of 1964[217] governing federal employment discrimination—an area of law now policed by a robust plaintiff's bar.

That approach, however, does not work for online fraud. Attorney's fees are typically unavailable for common-law fraud and state consumer-protection claims.[218] Moreover, employment-discrimination damages tend to be much larger than online consumer fraud claims, which typically amount only to a few hundred dollars.[219] Were attorney's fees available to the prevailing litigant, consumer plaintiffs could themselves be forced to pay for the defendant's attorney's fees should their claims fail. Thus, they would be unlikely to risk getting socked paying for both their own and their opponent's attorney's fees for so small a sum.

Second, the other common solution to the small-claims problem is the class action, which permits a class representative to assert a claim on behalf of not only herself but an entire class of individuals who suffered from the same wrong.[220] Private class action litigation has met with some notable successes in policing systematic, small-scale consumer deception.[221] But class action claims based on deceptive online ads and user-interface designs often fail to satisfy Federal Rules of Civil Procedure Rule 23's commonality requirement because apps and websites change so regularly:[222] Users in one region may be presented with different interface designs than users in another, and the designs may change from week to week. Indeed, that is a particular problem with targeted advertising techniques, which are specifically intended to present individually persuasive designs rather than designs of mass appeal. Private enforcement, even bolstered by the class-action procedure, will often be ineffective where the user experience varies widely between consumers.

---

[216] For detailed discussion of the so-called American rule, pursuant to which each party typically bears her own attorney's fees, see generally Peter Karsten, *Enabling the Poor to Have Their Day in Court: The Sanctioning of Contingency Fee Contracts, a History to 1940*, 47 DEPAUL L. REV. 231 (1998); Walter Olson & David Bernstein, *Loser-Pays: Where Next?*, 55 MD. L. REV. 1161 (1996); John Leubsdorf, *Toward a History of the American Rule on Attorney Fee Recovery*, 47 LAW & CONTEMP. PROBS. 9 (1984).

[217] Civil Rights Act of 1964, Pub. L. No. 88-352, 78 Stat. 241 (1964) (codified at 42 U.S.C. §§ 2000e–2000e-17).

[218] *See, e.g.*, Miller v. Argumaniz, 479 S.W.3d 306, 314 (Tex. App. 2015) ("Attorney fees are generally not recoverable for common law fraud.").

[219] *See supra* Part III.B.2.

[220] *See* FED. R. CIV. P. 23.

4. Calculated Arbitrators

Last are the *calculated arbitrators*. Historically, arbitration agreements were used primarily to resolve commercial disputes between businesses.[223] Arbitration offers a less expensive alternative to litigation, owing to its reduced formality; and it also provides greater privacy to its participants should the parties wish to keep the matter quiet. Over the last forty years, arbitration provisions have become increasingly popular, as the United States Supreme Court has broadened their availability by interpreting the Federal Arbitration Act[224] to require that state and federal courts "place arbitration agreements on an equal footing with other contracts" in all contexts and to refuse to enforce them only pursuant to "generally applicable contract defenses" and not "defenses that apply only to arbitration."[225]

Importantly, the Court has interpreted that requirement to apply with equal force to consumer contracts and to arbitration provisions that mandate individualized arbitration and forbid consumers from pursuing collective relief via class action.[226] Mandatory arbitration provisions now appear in all manner of consumer agreements, including for products and services sold online and in website and smartphone app terms of service agreements. One study estimates that consumer arbitration agreements apply to more than sixty percent of online sales in the United States.[227] Among large Fortune 100 companies, eighty-one rely on arbitration agreements in connection with consumer transactions, and seventy-eight use agreements that include class waivers that bar aggregation of claims by multiple consumers.[228]

---

[221] For a recent example, see Sellers v. JustAnswer LLC, 289 Cal. Rptr. 3d 1, 4–5 (Ct. App. 2021) (challenging website's practice of enrolling online consumers in automatic membership renewal option without adequate disclosure).

[222] *See* FED. R. CIV. P. 23(a)(2) (requiring "questions of law or fact common to the class"); Wal-Mart Stores, Inc. v. Dukes, 564 U.S. 338, 342 (2011) (applying Rule 23's commonality requirement).

[223] *See* BRIAN T. FITZPATRICK, THE CONSERVATIVE CASE FOR CLASS ACTIONS 13–16 (2019) (discussing the early history of the Federal Arbitration Act).

[224] Federal Arbitration Act, Pub. L. No. 68-401, 43 Stat. 883 (1925) (codified as amended at 9 U.S.C. §§ 1–402).

[225] AT&T Mobility LLC v. Concepcion, 563 U.S. 333, 339 (2011) (quoting Doctor's Assocs., Inc. v. Casarotto, 517 U.S. 681, 687 (1996)).

[226] *Id.* at 341–43.

Online services and apps include terms of service mandating individualized arbitration almost as a matter of course. Every time the user clicks, "I agree," on a website or downloads a new app, there is a high likelihood that she has agreed to arbitrate any dispute and waived the right to pursue a class action. The result is that in a large and rising number of online scams, a victim is obligated to pursue legal relief through private arbitration, not the court system. More importantly, because arbitration provisions typically disallow claim aggregation through class action, there is no way for consumers to combine their claims if, as is common, each is individually too small to justify legal action. Both legitimate companies and online scammers know this and have an incentive to include arbitration agreements for the purpose of discouraging even meritorious action against them.[229]

### C. A Coordinated Public-Private Response

We are now in a position to see the value of a coordinated public-private response to online deception. As explained above, private enforcement via state common law and unfair-competition statutes can be an extremely powerful tool to deter online fraud and enable victims to obtain redress.[230] Because such claims involve ex post application of broad standards against deceptive practices rather than ex ante proscription of dangerous technologies, innovative, but lawful, uses of those technologies can continue unimpeded.[231] And because they are brought by individual litigants, an enforcement regime centered around private actions can aggregate information from consumers across the nation and muster nearly unlimited enforcement resources from the private sector.[232]

Despite the partial deterrent effect of private enforcement, however, online deception has flourished by exploiting the legal patterns of deception, which allow scammers to avoid private litigation by concealing their identities, operating from foreign jurisdictions, employing rapidly changing user-interface designs that prevent class litigation, or extracting sums from large numbers of victims and imposing arbitration and class-action waiver provisions that preclude aggregate litigation.

---

[227] Imre Stephen Szalai, *The Prevalence of Consumer Arbitration Agreements by America's Top Companies*, U.C. DAVIS L. REV. ONLINE 233, 234 (2019), https://lawreview.sf.ucdavis.edu/sites/g/files/dgvnsk15026/files/media/documents/52-online-Szalai.pdf [https://perma.cc/HG5K-KQKL].
[228] *Id.*
[229] *See* Keith N. Hylton, *The Economics of Class Actions and Class Action Waivers*, 23 SUP. CT. ECON. REV. 305, 305–08 (2015) (analyzing circumstances in which companies will have incentives to include class waiver provisions in arbitration agreements).
[230] *See supra* Part III.A.
[231] *See supra* Part II.B.
[232] *See supra* Part III.A.

As I have argued elsewhere, one promising approach is federal or state legislation providing for statutory damages or attorney's fees to spur litigation challenging even small-sum deceptive practices.[233] For that approach to work, it might also be necessary to modify the Federal Arbitration Act to prohibit class-action waiver provisions or to prohibit mandatory arbitration of consumer claims altogether.[234]

Legislative action is not forthcoming on either and thus, this Section instead proposes what is a second-best approach: coordinated public-private enforcement within the existing legal regime. What we have for now is a resource-constrained FTC bolstered by better informed and better resourced private litigants who have some ability to protect their own interests, but whose efforts are hindered in many instances by procedural obstacles to private litigation. Faced with such a challenge, the best approach is for the FTC to deploy its limited enforcement resources carefully—to intentionally target those instances of online deception where the FTC has the greatest comparative advantage over private litigants, that is, where the Legal Patterns of Deception have prevented private enforcement.

Consider again the *fly-by-nighters*.[235] Cross-border scams now constitute more than ten percent of fraud reports to the FTC.[236] Scammers operating from foreign or unknown jurisdictions often makes private enforcement impossible or prohibitively expensive. Yet here the FTC Act and the U.S. SAFE WEB Act of 2006[237] give the FTC major advantages over private litigants. The SAFE WEB Act provides the FTC authority to pursue even foreign unfair or deceptive acts or practices so long as they are either "likely to cause reasonably foreseeable injury within the United States" or "involve material conduct occurring within the United States."[238] The act also empowers the FTC to provide and receive investigative assistance from foreign law-enforcement agencies, including foreign employee-exchange programs, where FTC staff are assigned to work on a case in a foreign jurisdiction or vice versa,[239] and by the sharing of confidential information obtained by compulsory process.[240] For example, the FTC can issue a civil investigative demand to compel a United States-based internet service provider to turn over information relevant to a cross-border fraud investigation, and then share the information with foreign law-enforcement officials to aid their own investigation of the target.[241] The FTC's major advantages over private litigation in policing cross-border fraud suggest that those cases should be a major focus of its enforcement efforts.

---

[233] *See* Dickinson, *supra* note 123, at 1665–67.

[234] *See, e.g.*, FITZPATRICK, *supra* note 223, at 125.

[235] *See supra* Part III.B.

[236] *See* FTC CROSS-BORDER FRAUD REPORT, *supra* note 198, at 1, 3–4.

[237] Undertaking Spam, Spyware, And Fraud Enforcement With Enforcers beyond Borders (U.S. SAFE WEB) Act of 2006, Pub. L. No. 109-455, 120 Stat. 3372 (codified as amended in scattered sections of 15 U.S.C. and 12 U.S.C. § 3412(e)).

[238] 15 U.S.C. § 45(a)(4)(A).

[239] *Id.* § 57c-1.

The FTC holds special advantages in combating *user-interface shapeshifters*[242] as well. First, under the FTC Act, the FTC is entitled to "prosecute any inquiry necessary to its duties"[243] and to "gather and compile information concerning . . . the organization, business, conduct, practices, and management of any person, partnership, or corporation . . . whose business affects commerce."[244] This gives the FTC the power to issue civil investigative demands to obtain information from entities it thinks may have violated the law or, just as importantly, from any other individual or entity who may have information related to the investigation. These powers go far beyond the typical civil litigation discovery process, where a litigant would be required to make out some plausible claim before proceeding to discovery and, even then, might be permitted to discover information only about the particular software products and interface design elements that were presented to her, not to other users. The FTC is thus in a position to successfully combat online deception even where quickly changing software products might impede aggregate litigation.

The FTC also has important advantages over private litigants in dealing with *nickel-and-dimers* and *calculated arbitrators*. The primary obstacle to private litigation against such scammers, recall, is the limited recovery available for small losses combined with some obstacle to claim aggregation—for example, harms are too diverse to permit class certification or a class action waiver clause in an arbitration agreement.[245] Fortunately, the FTC is again well positioned to overcome these concerns. Unlike private litigants, government-funded FTC attorneys can and will bring enforcement actions even where potential recovery is small. Moreover, the FTC is of course not bound by any arbitration agreement entered into by a consumer. And because the agency al-ready represents every one of the nation's consumers, there is no need for it to seek and obtain class certification before seeking widespread relief.

---

[240] *Id.* § 57b-2(b)(6).
[241] *See An Explanation of the Provisions of the US SAFE WEB Act* 2–4, FED. TRADE COMM'N, https://www.ftc.gov/sites/default/files/documents/reports/us-safe-web-act-protecting-consumers-spam-spyware-and-fraud-legislative-recommendation-congress/explanation-provisions-us-safe-web-act.pdf [https://perma.cc/83JY-AZUN] (explaining the U.S. SAFE WEB Act's changes to allow the FTC to share information with foreign entities and aid investigations and enforcement actions).
[242] *See supra* Part III.B.3.
[243] 15 U.S.C. § 43.
[244] *Id.* § 46(a).
[245] *See supra* Part III.B.

The comparative advantages of FTC enforcement suggest that it is well suited to police online deception even in those circumstances where procedural obstacles have prevented private litigants from doing so. That observation has important and counterintuitive implications for how the FTC should allocate its very limited resources. First, the FTC should favor enforcement actions against foreign, rather than domestic scammers. A significant number of online scams are operated by foreign entities, against whom private litigation is impossible or unduly expensive for any but the very largest losses. Second, the FTC should focus its efforts disproportionately on small-scale but widespread schemes, particularly those by entities whose consumer contracts include arbitration clauses with class-action waivers. Once stripped of the power to aggregate their claims, private litigants are unable to effectively prosecute such minor claims. Finally, the FTC will be far more effective than private enforcement in the context of hyper-targeted consumer advertising campaigns, frequently updated smartphone apps, and other quickly evolving products and services, whose rapid changes may preclude certification of a private class.

## Conclusion

As digital tools continue to drive an explosion in online fraud, the law's response has been only half right. Modern online scams *are* different, but what is different about them is not, primarily, any new power to deceive, but their unprecedented efficiency, which has driven growth in online scams at a rate that far outpaces the FTC's enforcement capacity. That observation has major implications for lawmakers, whose well-meaning efforts to identify and restrict technologies driving deception are both unnecessary and counterproductive. More important than the technologies that power deception, and more amenable to regulatory action, are the legal obstacles to private enforcement that have allowed online scams to flourish. A superior approach would coordinate public and private enforcement efforts by targeting governmental resources at combatting the recurring patterns of deception that help scammers to evade justice, thereby more effectively combating online fraud while also avoiding new impediments to technological innovation.